\documentclass{aa}

\usepackage{multirow}
\usepackage{graphicx}
\usepackage{verbatim}
\usepackage{placeins}
\usepackage{txfonts}
\usepackage[dvipsnames]{xcolor}
\usepackage{hyperref}
\hypersetup{
    colorlinks=true,
    linkcolor=blue,
    filecolor=magenta,
    urlcolor=blue,
    citecolor=blue
}

\providecommand{\plato}{\textsc{Plato}}
\providecommand{\gaia}{\textit{Gaia}}

\newcommand{\teff}{T_{\mathrm{eff}}} 
\newcommand{\lgg}{\log{g}} 
\newcommand{\lggu}{\log{g / \mathrm{cm\,s^{-2}}}} 
\newcommand{\vsini}{\varv\,\sin i} 
\newcommand{\vmic}{V_{\mathrm{turb}}} 
\newcommand{\feh}{\mathrm{\left[Fe/H\right]}} 
\newcommand{\afe}{\mathrm{\left[\alpha/Fe\right]}} 
\newcommand{\kms}{\mathrm{km\,s^{-1}}} 

\providecommand{\ag}{\ensuremath{A_G}\xspace}

\providecommand{\bpminrp}{\ensuremath{G_\mathrm{BP}-G_\mathrm{RP}}\xspace}
\providecommand{\gmag}{\ensuremath{G}}
\providecommand{\bpmag}{\ensuremath{G_\mathrm{BP}}}
\providecommand{\rpmag}{\ensuremath{G_\mathrm{RP}}}

\providecommand{\diam}{$\theta_{\rm LD}$}

\makeatletter 
\renewcommand*\aa@pageof{, page \thepage{} of \pageref*{LastPage}}\makeatother 

\begin{document}

   \title{Performance of the \textit{Stellar Abundances and atmospheric Parameters Pipeline} adapted for M~dwarfs}

   \subtitle{I. Atmospheric parameters from the spectroscopic module}

   \author{
   Terese Olander\inst{\ref{inst:UAO}}
   \and Matthew R. Gent\inst{\ref{inst:MPIA},\ref{inst:IRAP}}
   \and Ulrike Heiter\inst{\ref{inst:UAO}}
   \and Oleg Kochukhov\inst{\ref{inst:UAO}}
   \and Maria Bergemann\inst{\ref{inst:MPIA}}
   \and Ekaterina Magg\inst{\ref{inst:MPIA}}
   \and Santi Cassisi\inst{\ref{inst:INAF}}
   \and Mikhail Kovalev\inst{\ref{inst:Kovalev}}
   \and Thierry Morel\inst{\ref{inst:Liege}}
   \and Nicola J. Miller\inst{\ref{inst:UAO}}
   \and Diogo Souto\inst{\ref{inst:Sergipe}}
   \and Yutong Shan\inst{\ref{inst:Oslo},\ref{inst:Göttingen}}
   \and B\'arbara Rojas-Ayala\inst{\ref{inst:Tarapaca}}
   \and Elisa Delgado-Mena\inst{\ref{inst:Porto},\ref{inst:madrid}}
   \and Haiyang S. Wang\inst{\ref{inst:Copenhagen}}
   }

   \institute{
   Observational Astrophysics, Department of Physics and Astronomy, Uppsala University, Box 516, SE-751 20 Uppsala, Sweden, \email{terese.olander@physics.uu.se}, \email{ulrike.heiter@physics.uu.se},
   \label{inst:UAO}
   \and
   Max Planck Institute for Astronomy, 69117 Heidelberg, Germany
   \label{inst:MPIA}
   \and
   Institut de Recherche en Astrophysique et Planétologie, Université de Toulouse, CNRS, IRAP/UMR 5277, 14 Avenue Edouard Belin, 31400, Toulouse, France, \email{mgent@irap.omp.eu}
   \label{inst:IRAP}
   \and
   INAF-Osservatorio Astronomico d'Abruzzo, Via M. Maggini, s/n, 64100, Teramo, Italy; INFN, Sezione di Pisa, Largo Pontecorvo 3, 56127 Pisa, Italy
   \label{inst:INAF}
   \and
   Yunnan Observatories, China Academy of Sciences, Kunming 650216, China; Key Laboratory for the Structure and Evolution of Celestial Objects, Chinese Academy of Sciences, Kunming 650011, China
   \label{inst:Kovalev}
   \and
   Space sciences, Technologies and Astrophysics Research (STAR) Institute, Université de Liège, Quartier Agora, Allée du 6 Août 19c, Bât. B5c, B4000 Liège, Belgium
   \label{inst:Liege}
   \and
   Departamento de F\'isica, Universidade Federal de Sergipe, Av. Marcelo Deda Chagas, S/N Cep 49.107-230, S\~ao Crist\'ov\~ao, SE, Brazil
   \label{inst:Sergipe}
   \and
   Centre for Planetary Habitability, Department of Geosciences, University of Oslo, Sem Sælands vei 2b, 0315 Oslo, Norway
   \label{inst:Oslo}
   \and
   Institut für Astrophysik, Georg-August-Universität, Friedrich-Hund-Platz 1, 37077 Göttingen, Germany
   \label{inst:Göttingen}
   \and
   Instituto de Alta Investigaci\'on, Universidad de Tarapac\'a, Casilla 7D, Arica, Chile
   \label{inst:Tarapaca}
   \and
   Instituto de Astrofísica e Ciências do Espaço (IA), CAUP, Universidade do Porto, Rua das Estrelas, PT4150-762 Porto, Portugal
   \label{inst:Porto}
   \and
      Centro de Astrobiolog\'ia (CAB), CSIC-INTA, Camino Bajo del Castillo
s/n, 28692, Villanueva de la Ca\~nada (Madrid), Spain
\label{inst:madrid}
   \and
   Center for Star and Planet Formation, Globe Institute, the University of Copenhagen, Øster Voldgade 5-7, 1350 København K, Denmark
   \label{inst:Copenhagen}
   }

   \date{Received ; accepted }


  \abstract
   {M~dwarfs are important targets in the search for Earth-like exoplanets due to their small masses and low luminosities. Several ongoing and upcoming space missions are targeting M~dwarfs for this reason, and the ESA \plato\ mission is one of these.}
   {In order to fully characterise a planetary system the properties of the host star must be known. For M~dwarfs we can derive effective temperature, surface gravity, metallicity, and abundances of various elements from spectroscopic observations in combination with photometric data.}
   {The Stellar Abundances and atmospheric Parameters Pipeline (SAPP) has been developed to serve as a prototype for one of the stellar science softwares within the \plato\ consortium. The pipeline combines results from a spectroscopy, a photometry, an interferometry, and an asteroseismology module to derive stellar parameters for FGK-type stars. We have modified the pipeline to be able to analyse the M~dwarf part of the \plato\ target sample. The current version of the pipeline for M~dwarfs mostly relies on spectroscopic observations. The module processing these data is based on the machine learning algorithm The Payne and fits a grid of model spectra to an observed spectrum to derive effective temperature and metallicity. We use spectra in the H-band, as the near-infrared region is beneficial for M~dwarfs because there are fewer molecular lines and they are brighter in this wavelength region than in the optical. A method based on synthetic spectra was developed for the continuum normalisation of the spectra, taking into account the pseudo-continuum formed by numerous lines of the water molecule. Photometry is used to constrain the surface gravity.}
   {We tested the modified SAPP on spectra of M~dwarfs from the APOGEE survey. Our validation sample of 26 stars includes stars with interferometric observations and binaries. We found a good agreement between our derived values and reference values from a range of previous studies. We estimate the overall uncertainties in the derived effective temperature, surface gravity, and metallicity to be 100~K, 0.1~dex, and 0.15~dex, respectively.}
   {We find that the modified SAPP performs well on M~dwarfs and identify possible areas of future development that should lead to an improved precision of the derived stellar parameters.}

   \keywords{techniques: miscellaneous -- stars: fundamental parameters -- stars: atmospheres -- stars: late-type -- stars: low-mass -- infrared: stars}

   \maketitle


\section{Introduction}
\label{sec:introduction}
M~dwarfs have become prime targets in the hunt for terrestrial exoplanets. This is partly due to their multitude, as over 70~\% of the stars in the solar neighbourhood are estimated to be M~dwarfs \citep{Henry2006NrMdwarfs}. In addition, their low luminosities and small radii make it easier to find planets around them using both the radial velocity and the transit methods. Observing transits of exoplanets in the habitable zone around M~dwarfs and obtaining high resolution transmission spectra of the planetary atmosphere is also made easier by the shorter orbital periods due to the proximity of the exoplanet to the star. Examples for recent exoplanet transmission spectroscopy observations for M~dwarfs can be found in \citet{2023AJ....165..170R, 2023AJ....165..169D, 2022AJ....164..225D}. 

However, a key challenge to understanding M-dwarf planets lies in the difficulty of spectroscopically characterising the stars themselves. Being at the faint end of the main sequence, M~dwarfs have effective temperatures below 4000~K.
These low temperatures make it possible for molecules to form, and the photosphere of M~dwarfs therefore has many di- and triatomic molecules.
The optical wavelength region is dominated by bands of TiO, which hide the atomic lines, and shows spectral features of CaH, MgH, CaOH, and VO \citep[e.g.][]{GrayCorbally+2009+339+387,1995A&A...295..736B,2000ApJ...540.1005A,2016PhR...663....1S}. 
At longer wavelengths in the near-infrared (NIR) there are some regions with fewer molecular lines but the atomic lines are still severely blended with lines from molecules such as CO, FeH, OH, and H$_2$O \citep[e.g.][]{GrayCorbally+2009+339+387,2012ApJ...748...93R,Lind2016,2022ApJSouto}. 
It is however easier to distinguish the atomic lines at these longer wavelengths, since the absorption features of these molecules are not as dense and deep as those of TiO. The plethora of lines, both atomic and molecular, complicates the analysis of M~dwarf spectra. This applies in particular to the continuum normalisation process, as it becomes increasingly difficult to identify the continuum for decreasing effective temperature.
For example, in the H-band the presence of water molecules suppresses the continuum to form what we call a pseudo-continuum. This has to be taken into account when normalising an observed spectrum (see Sect.~\ref{subsubsec:method:normalisation} and \citealt{APOGEE_Sarmento2021}).

Ideally, spectroscopically derived parameters should be verified by comparing with parameters obtained with independent methods based on measurements of fundamental stellar properties. Examples are effective temperatures derived from interferometric angular diameters and surface gravity derived from asteroseismology.
However, asteroseismology cannot be used for M~dwarfs because their pulsations cannot be detected with current observing capabilities \citep{Rodriguez-Lopez2019PulsMdwarf}.
Interferometry also has some challenges in regards to M~dwarfs. The angular diameter is obtained via observations, and then the Stefan-Boltzmann law is used together with the bolometric flux of the star to derive the temperature. At the moment this method can only be used on few M~dwarfs, because of the faintness and small radius of these stars\footnote{The magnitude limit of current instruments is at about $H$=7, corresponding to $V$ around 11 for M~dwarfs. The limit for angular diameter lies at 0.3~mas in the best conditions, corresponding to stellar radii of $\sim$0.3~R$_\odot$ at a distance of 10~pc, $\sim$0.6~R$_\odot$ at 20~pc, or $\sim$1~R$_\odot$ at 30~pc \citep[e.g.,][]{Boyajian2012,2019MNRAS.484.2656L,2022SPIE12183E..08M}.}. 
\citet{Boyajian2012} and \citet{Rabus2019} give effective temperatures, angular diameters and surface gravities or masses for a range of M~dwarfs, see Sect.~\ref{subsec:result:comparison:interferometry}.

Despite these challenges, many spectroscopic studies determining stellar parameters of M~dwarfs have presented promising results in recent years.
Classical tools for spectrum synthesis and fitting have been applied to low-resolution optical SNIFS \citep{2004SPIE.5249..146L} and NIR SpeX \citep{2003PASP..115..362R} spectra calibrated to an absolute flux scale \citep{2015ApJMann},
high-resolution optical HARPS and HARPS-N spectra smoothed to low resolution
\citep{2015A&A...577A.132M,2020AAMaldonado},
high-resolution optical Keck-HIRES spectra \citep{2021ApJS..255....8R}, 
as well as high-resolution NIR spectra, including
spectra in the J~band from the CRIRES spectrograph \citep{Lind2016,Lind2017},
spectra from the CARMENES spectrograph \citep[optical plus J- and H~bands,][]{Pass2018,Pass2019,Rajpurohit2018CARMENES,2021A&A...656A.162M}, 
spectra from the APOGEE survey \citep[H~band,][]{APOGEE_Sarmento2021,2022ApJSouto,2024ApJ...973...90M}, 
and spectra from the SPIRou spectrograph \citep[J\nobreakdash-, H-, and K~bands,][]{2022MNRAS.516.3802C,2022MNRAS.511.1893C}. 
At the same time, machine learning approaches have started to be used for the analysis of optical and NIR spectra, including
\citet[][optical spectra from five instruments]{2020A&A...636A...9A,2024A&A...690A..58A}, 
\citet[][optical WiFeS spectra]{2024MNRAS.529.3171R}, 
\citet{2020A&A...642A..22P,2022AaA...658A.194P}, \citet{2023A&A...673A.105B}, and \citet{2024A&A...687A.205M} for CARMENES, 
as well as \citet{The_Payne_Ting2019ApJ} and \citet{The_Cannon_Mdwarf_Birky2020ApJ} for APOGEE.
For details on some of these studies, which we used for comparison, see Sect.~\ref{subsec:result:comparison:spectroscopy}.




Current and upcoming large surveys that include M~dwarfs in their samples put new demands on deriving accurate parameters and abundances. One such project is \plato\ \citep{2024arXiv240605447R}, 
a space telescope that is planned to be launched in 2026. Its mission is to find habitable terrestrial exoplanets around solar type stars. In addition, at least 5000 (possibly up to about 25\,000) M~dwarfs with a $V$ magnitude brighter than 16 are planned to be part of the sample. Additional selection criteria are applied for the M~dwarf sample, based on \gaia\ \bpminrp\ colours and absolute \gmag\ magnitudes \citep{2021A&A...653A..98M,2022A&A...658A..31N}. 
The $V$ magnitude selection criterion leads to the \plato\ sample to be dominated by early- to mid M~dwarfs with very few targets with spectral types later than M4 expected to be retained.
\plato\ will obtain light curves in order to detect exoplanets and to apply asteroseismology to the host stars (those of FGK-type) in order to derive stellar parameters. There will also be a spectroscopic follow-up of the targets.
This requires a fast and reliable method to analyse the spectra of thousands of stars.

In preparation of the mission, the Stellar Abundances and atmospheric Parameters Pipeline \citep[SAPP,][]{Gent_2022A&A} has been developed to serve as a prototype for one of the components of the \plato\ stellar science software. 
This pipeline uses Bayesian inference to obtain accurate stellar parameters such as $\teff$, $\lgg$, $\feh$, and chemical abundances from spectroscopic, photometric, and asteroseismic data. The SAPP has so far only been tested on FGK stars in the optical wavelength region \citep{Gent_2022A&A}. Here we present the results from a modified version of the SAPP capable of analysing M~dwarfs in the H-band (see Sect.~\ref{sec:Method}), which does not use the asteroseismic and the Bayesian inference part of the code. Our modified SAPP pipeline uses high-resolution spectroscopic and photometric data to derive effective temperature, surface gravity, and metallicity. We leave the analysis of abundances and the full Bayesian analysis to future work.

In Sect.~\ref{sec:Sample} we present the observed sample we used in order to test the pipeline. In Sect.~\ref{sec:Method} we present the pipeline and its different components, including the sources for the photometric input data needed in the analysis. We give the results and compare with literature values for the sample stars in Sect.~\ref{sec:Results}. We end with an outlook and conclusions in Sects.~\ref{sec:future} and \ref{sec:conclusion}, respectively.


\section{Sample and spectroscopic data}
\label{sec:Sample}

\begin{table*}[ht]
       \centering
       \caption{Stars in our sample.}
              \label{tab:Sample}
        \begin{tabular}{lllrlcrllr}
        \noalign{\smallskip}
        \hline
        \noalign{\smallskip}
        Star &   RA [J2000] &  DEC [J2000] &  $K_s$ mag & Sp. type & Ref. & $\vsini$ [$\kms$] & Int. & Bin. & S/N \\
        \hline
        \noalign{\smallskip}
      BD+00549B & 03 15 00.9229 & +01 03 08.186 & 10.855 & M1.0  & 1  & $<8.0$ &      & FGK/M & 120 \\
     BD-064756B & 18 24 46.8871 & -06 20 31.359 &  8.795 & M4    & 2  &  &      & FGK/M & 580 \\
         GJ105A & 02 36 04.9013 & +06 53 12.383 &  3.520 & K3    & 3  &  & B    & K/M   & 230 \\
         GJ105B & 02 36 15.2668 & +06 52 17.915 &  6.574 & M3.5  & 4  &  &      & K/M   & 185 \\
          GJ15A & 00 18 22.8849 & +44 01 22.637 &  4.020 & M2    & 3  &  & B    &       & 125 \\
          GJ205 & 05 31 27.3957 & -03 40 38.024 &  3.900 & M1.5  & 4  & 2.1 & B, R &       & 195 \\
          GJ212 & 05 41 30.7306 & +53 29 23.290 &  5.759 & M1.0  & 4  & 1.9 &      & FGK/M & 110 \\
       GJ297.2B & 08 10 34.2940 & -13 48 51.131 &  7.418 & M2.5  & 4  & $<8.0$ &      & FGK/M & 550 \\
         GJ3195 & 03 04 43.4430 & +61 44 08.853 &  8.103 & M2.5  & 4  & $<8.0$ &      & FGK/M & 195 \\
         GJ324B & 08 52 40.8627 & +28 18 58.821 &  7.666 & M4.5  & 4  & 0.0 &      & FGK/M & 770 \\
         GJ338A & 09 14 22.7748 & +52 41 11.791 &  3.990 & K7    & 5  & 1.0 & B    & K/M   & 140 \\
         GJ338B & 09 14 24.6828 & +52 41 10.902 &  4.140 & M0    & 5  & 0.5 & B    & K/M   & 150 \\
          GJ393 & 10 28 55.5512 & +00 50 27.598 &  5.311 & M2    & 6  & 0.0 &      &       & 160 \\
         GJ412A & 11 05 28.5769 & +43 31 36.386 &  4.769 & M1.0  & 7  & 0.0 & B    &       & 140 \\
          GJ447 & 11 47 44.3972 & +00 48 16.400 &  5.654 & M4    & 6  & 0.0 & R    &       & 230 \\
          GJ526 & 13 45 43.7755 & +14 53 29.471 &  4.415 & M2    & 3  & 0.0 & B, R &       & 140 \\
          GJ687 & 17 36 25.8993 & +68 20 20.909 &  4.548 & M3.0  & 4  & 0.0 & B, R &       &  40 \\
         GJ725A & 18 42 46.7043 & +59 37 49.409 &  4.432 & M3    & 5  & 1.0 & B    & M/M   & 125 \\
         GJ725B & 18 42 46.8946 & +59 37 36.721 &  5.000 & M3.5  & 5  & 1.4 & B    & M/M   &  90 \\
         GJ752A & 19 16 55.2565 & +05 10 08.040 &  4.673 & M3-   & 3  & 0.2 &      &       &  50 \\
         GJ777B & 20 03 26.5810 & +29 51 59.529 &  8.712 & M4.5  & 4  & $<8.0$ &      & FGK/M &  80 \\
          GJ809 & 20 53 19.7889 & +62 09 15.817 &  4.618 & M1.0  & 7  & 0.0 & B, R &       &  30 \\
          GJ880 & 22 56 34.8046 & +16 33 12.355 &  4.523 & M1.5  & 7  & 1.3 & B, R &       & 130 \\
LSPMJ0355+5214  & 03 55 36.8973 & +52 14 28.967 & 10.127 & M2.5  & 4  & $<8.0$ &      & FGK/M & 300 \\
LSPMJ1204+1728S & 12 04 56.1109 & +17 28 11.434 &  8.967 & M3.5  & 4  & 17 &      & FGK/M & 405 \\
        Ross799 & 14 04 55.8381 & +01 57 23.085 &  9.269 & M2    & 8  & $<8.0$ &      & FGK/M & 305 \\
        \noalign{\smallskip}
        \hline
        \end{tabular}
        \tablefoot{GJ~105A is mid-K~dwarf whose literature parameters put it in the validity range of our pipeline. GJ~338A is a very late K or an early M~dwarf. $K_s$ magnitudes are from 2MASS \citep{2003Cutri}. Column ``Ref.'' gives the reference for the spectral type. In the spectral type range spanned by our sample, \citet{1989ApJS...71..245K} assigned subclasses K3, K4, K5, M0, M1, M2, M3, M4, with a sub-division to a quarter of a subclass (e.g., M0, M0+, M0.5, M1-, M1). Values of $\vsini$ are taken from \citet[based on CARMENES spectra]{Reiners2022} and \citet[upper limits and LSPM~J1204+1728S value, based on APOGEE spectra]{2018AJGilhool_Mdwarf_rot_APOGEE}. Column ``Int.'' indicates that an interferometric value of $\teff$ is available from \citet[][B]{Boyajian2012} and/or \citet[][R]{Rabus2019}. Column ``Bin.'' indicates that the star is in a binary with an FGK primary (FGK/M), where the secondary was analysed with the modified SAPP, or with a late K or M primary (K/M or M/M), for which both components were analysed with the SAPP (see Sect.~\ref{subsec:comp_bin}). The S/N is the flux of the APOGEE spectra divided by its uncertainty, averaged over the whole wavelength range, and rounded to the nearest multiple of 5.}
        \tablebib{(1) \citet{2013AJ....145...52M}; (2) \citet{2019ApJ...877...60B}; (3) \citet{1989ApJS...71..245K}; (4) \citet{2015A&A...577A.128A}; (5) \citet{1991ApJS...77..417K}; (6) \citet{2019AJ....157...63K}; (7) \citet{2013AJ....145..102L}; (8) \citet{1985ApJS...59..197B}}
   \end{table*}


We tested the M~dwarf version of the SAPP on observed spectra from the APOGEE survey \citep{2017AJ....154...94M,2020AJ....160..120J}, 
which is part of the Sloan Digital Sky Survey \citep[SDSS-IV,][]{2017AJ....154...28B}. 
The targets of the survey are primarily red giants, but M~dwarfs are among the observed stars.
We chose APOGEE due to the large number (order of thousand) of reduced M~dwarf spectra available with a resolution sufficient for \plato's precision needs \citep{APOGEE_Sarmento2021}.
We note that a considerable number (over 300) of reduced CARMENES near-infrared spectra have become available recently \citep{2023A&A...670A.139R}. 
A comparative study will be carried out in future work, following this work and \citet{2022AaA...658A.194P}. 

Our test sample was chosen to cover a range from early- to mid-M~dwarfs, consistent with the properties of the \plato\ target stars (see Sect.~\ref{sec:introduction}) and consists of 26 stars. Two K~dwarfs with APOGEE spectra were included in this sample as their literature stellar parameters are within our parameter limits (see Sect.~\ref{subsubsec:method:spectra}) and they are in a binary with an M~dwarf that is also included in the sample.

All stars in the sample have determinations of atmospheric parameters in the literature.
Stars with interferometric effective temperatures as well as M~dwarfs in wide binaries with other M~dwarfs and with FGK~stars are included in our sample. Interferometric measurements give model independent temperatures, and together with reliable distances and mass-luminosity relations surface gravities can be obtained. Binaries are excellent benchmark systems for verifying metallicities and chemical abundances, assuming that the component stars have formed from the same material within a molecular cloud and thus have the same chemical composition \citep[e.g.][]{2004A&A...420..683D,2006A&A...454..581D}.
The literature data are described in detail in Sect.~\ref{sec:Results}.

The APOGEE survey makes use of two multi-object spectrographs \citep{2019PASP..131e5001W}; APOGEE-N on the Sloan 2.5~m telescope in New Mexico \citep{2006AJ....131.2332G} and APOGEE-S on the 2.5~m duPont telescope in Chile \citep{1973ApOpt..12.1430B}. 
APOGEE has a resolving power of $R=\lambda/\Delta\lambda \sim$22\,500 and covers the H-band ($15~000$ to $17~000$~\AA).
Three Hawaii-2RG detectors are used, where each detector covers about a third of the wavelength range.
We used the combined spectra from multiple observations (\textit{apStar/asStar} files) available from SDSS-IV DR14 and DR16\footnote{\url{https://dr14.sdss.org/home}, \url{https://dr16.sdss.org/home}, \url{https://live-sdss4org-dr16.pantheonsite.io/irspec/spectral_combination/}}.
The observed spectra were reduced by the APOGEE pipeline \textit{apred} \citep{2015AJ....150..173N}. The spectra are wavelength calibrated and have had telluric lines removed. The spectra are also radial velocity (RV) corrected. However, we found some discrepancy for some of the stars in our sample, so the SAPP was used to recalculate the RV shift for all stars in the sample.
We refer the reader to \citet{Gent_2022A&A} for more details on the RV correction.

The entire sample can be seen in Table~\ref{tab:Sample} together with the coordinates, 2MASS $K_s$ magnitude, spectral type, projected equatorial rotational velocity, availability of interferometric data, binary system specification, and S/N. The S/N is between 100 and 400 for most stars, with a few exceptions towards lower and higher values.


\section{Method}
\label{sec:Method}
The SAPP serves as a prototype for one of the components of the stellar science software that will derive stellar parameters and abundances for stars in the \plato\ sample. This includes both FGK stars and M~dwarfs.
The original FGK version of the SAPP consists of modules based on spectroscopy, photometry, and interferometry, together with asteroseismology. In its full version the code uses Bayesian inference on results from the different modules to derive parameters such as $\teff$, $\lgg$, $\feh$, and chemical abundances. For a thorough description of all the functions the reader is directed to \citet{Gent_2022A&A}.

The M-dwarf version of the SAPP presents several key differences. Due to the unobservable nature of M~dwarf pulsations the asteroseismic module cannot be applied. 
Adopting a $\lgg$ determined from the granulation properties \citep{2018A&A...620A..38B} also appears unlikely.
In addition, the original SAPP operates on optical spectra, which for M~dwarfs are blanketed by many molecular lines. We therefore modified the spectroscopic module such that NIR spectra in the H~band are analysed within the parameter range of M~dwarfs.
Furthermore, the spectroscopic analysis of M~dwarfs is prone to degeneracies, a well-known one being the $\teff$--$\feh$ degeneracy \citep[e.g.,][]{Pass2018}.
\citet{Pass2018} showed that the use of an independent method to constrain $\lgg$ helps to reduce this degeneracy. Therefore, we derive $\lgg$ from photometry and stellar evolution models, before using it to determine $\teff$ and $\feh$ from spectroscopy.
It is similar to the pipeline for FGK stars where an external constraint on $\lgg$ is also used \citep{Gent_2022A&A}, as routinely done for asteroseismic targets nowadays \citep[e.g.][]{2024arXiv240515919L}. In this article we focus on obtaining reliable parameters using the spectroscopic module together with constraints from photometry, while the full Bayesian inference analysis including chemical abundances is left to future developments.
In this section we describe the modified version of the SAPP capable of analysing M~dwarf spectra.


\subsection{Model isochrones and photometry}
\label{subsec:method:photometry and evolution}

The photometric module is used to estimate fundamental stellar parameters by fitting model isochrones to broadband photometric data.
The stellar evolution models for M~dwarfs adopted in the present work were specifically calculated for the analysis of M~dwarfs among the \plato\ targets. They represent an extension of the set of models for very low-mass (VLM) stars presented in \citet{Hidalgo:2018} and \citet{pietrinferni:2021} in the framework of the updated BaSTI library\footnote{\url{http://basti-iac.oa-abruzzo.inaf.it}}. For a detailed discussion of the input physics and numerical assumptions adopted in performing the evolutionary computations, we refer the interested reader to the mentioned references. Here we only briefly summarise the input physics more relevant for the computations of the M~dwarf models adopted in present work. The adopted solar metal mixture is that provided by \cite{caffau:2011}, supplemented by the abundances given by \cite{lodders:2010}, see Table~1 in \citet{Hidalgo:2018}. By adopting this metal mixture, the calibration of the Solar Standard Model (SSM) provides the initial chemical composition for the Sun as $Z_{ini}=0.01721$ and $Y_{ini}=0.2695$, while the actual surface metallicity of the Sun results to be equal to $Z_\odot=0.0153$.

Superadiabatic convection in the outer layers is treated according to the \citet{bohm-vitense} flavour of the mixing length theory (MLT). The value of the free mixing length parameter $\alpha_{ml}$ was fixed to 2.006 by the SSM calibration. We note that in any case the calibration of the MLT is not an issue in the VLM stellar regime, as these stars are largely adiabatic along their whole interior structure. On the other hand, a crucial issue concerning M~dwarf stellar models is the treatment of the outer boundary conditions, as extensively discussed by \citet[][and references therein]{baraffe:1995,Brocato:1998,cb:00,cs:13}: for structures with a mass lower than about $0.45~M_\odot$ it is crucial to determine the outer boundary conditions via accurate non-gray model atmospheres in order to retrieve reliable and precise evolutionary predictions. For the present models we adopted the outer boundary conditions provided by the PHOENIX model atmosphere repository.
\citep{allard:12,2013Husser_PHOENIX}. For more details on this topic we refer to the discussion in \citet{Hidalgo:2018}.

The thermodynamical properties were obtained by using the FreeEOS equation of state by A. Irwin \citep{csi:2003,Hidalgo:2018}, in the configuration that provides the most accurate predictions in the thermal regime of high density and low temperature suitable for VLM stars. The sources for the radiative Rosseland opacity are the same as for the more massive stellar structures in the BaSTI library: opacities were taken from the OPAL calculations \citep{iglesias:96} for temperatures larger than log($T$) = 4.0, whereas for lower temperatures the predictions provided by \citet{ferguson:05} were adopted. The adopted conductive opacities were taken from the tabulations given by \citet{cassisi:2007,cassisi:21}.

The photometric module uses the same set of photometric bands as in \citet{Gent_2022A&A}, that is, Johnson $B$ and $V$ \citep{2010MNRASKoen,2003AJMonet,2012Zacharias}, \gaia\ \gmag, \bpmag, \rpmag\ \citep{2016A&A...595A...1G,2023A&A...674A...1G}, and 2MASS $J$, $H$, $K_s$ magnitudes \citep{2003Cutri}.
Future updates of the SAPP for M~dwarfs will likely also incorporate further NIR bands, such as W1 and W2 at 3.4 and 4.6~$\mu$m from the Wide-field Infrared Survey Explorer (WISE, \citealt{2021ApJS..253....8M}), 
to better capture the peak of the spectral energy distribution (SED) of M dwarfs.
The photometry is combined with photogeometric distances derived by \citet{2021AJBailar-Jones}. If distances are not available in that source, \gaia\ parallaxes \citep{2023A&A...674A...1G} are used by the pipeline to calculate the distances.
The \texttt{Stilism} tool \citep{Capitanio2017} was used to obtain line-of-sight reddening corresponding to the given distances, in order to derive and correct for interstellar extinction (see below). We note that the M~dwarfs in our sample are all nearby, and their reddening values are consistent with zero.

We computed a grid of model isochrones with synthetic photometry in the bands described above. The grid spans ages between 0.5~Gyr and 15~Gyr in steps of 0.5~Gyr, masses from 0.1 to 0.75~M$_\odot$ with steps of 0.005~M$_\odot$, and $\feh$ from $-2.45$ to +0.14~dex (+0.28~dex for the 0.5~Gyr isochrone) with steps of 0.01~dex. We chose a relatively coarse age grid because the stability of M dwarfs throughout their long main-sequence lifetimes means the age sensitivity of the isochrones is small.
This range accounts for changes significant in $\lgg$, specifically radius inflation for larger masses. The isochrones will be extended to higher [Fe/H] in future updates of the pipeline.

The synthetic photometry from the isochrones is then compared to observed absolute magnitudes using reddening E(B$-$V) and distance $d$ to derive extinction in all available photometric bands. Except for \gaia\ bands, the extinction 
is derived using $R$ values adopted from \cite{2011A&A...530A.138C}. 
If the extinction \ag from \gaia\ DR3 is not available, 
\gaia\ \bpminrp-colour-dependent coefficients presented in \cite{2021MNRAS.507.2684C}  
are used to derive the extinction for \gmag, \bpmag, and \rpmag.
By comparing model 
to observed absolute magnitudes, probability distribution functions (PDFs) are derived for the stellar parameters spanned by the BaSTI isochrones, that is, log($\teff$), $\lgg$, [Fe/H], log(mass/M$_\odot$), log(luminosity/L$_\odot$), log(radius/R$_\odot$) for all available ages.
See Sect.~3.3 in \citet{Gent_2022A&A} for the exact formulation for deriving the photometric PDF and the error propagation.

In \cite{Gent_2022A&A} a set of stellar parameters is passed to the photometric module to define a subdomain within the parameter space, minimising the number of stellar photometric models that are processed in the code. However, here we define the subdomain of the M~dwarf grid as 3500 $\pm$ 800~K, 4.9 $\pm$ 1.0~dex, and $-0.02$ $\pm$ 1.0~dex in $\teff$, $\lgg$, and $\feh$, respectively, regardless of the properties of the star being analysed, and the photometric module is only used to determine the $\lgg$.


\subsection{Spectroscopy}
\label{subsec:method:spectroscopy}

The spectroscopic module in the SAPP is based on The Payne,
a method developed by \citet{The_Payne_Ting2019ApJ} to infer stellar parameters from observed spectra based on a machine learning algorithm. The Payne uses an artificial neural network (ANN) model trained on a grid of synthetic spectra for FGKM dwarf and giant stars. \citet{The_Payne_Ting2019ApJ} generated the spectra with ATLAS12 model atmospheres and the SYNTHE spectral synthesis code. The spectra were defined by 25 parameters (or 'labels') corresponding to stellar properties, including $\teff$, $\lgg$, turbulence parameters, as well as elemental abundances. The code identifies the set of parameter values that best reproduce the observed spectrum.
\citet{The_Payne_Ting2019ApJ} tested The Payne on stars in the APOGEE DR14 sample, which includes M~dwarfs.
For FGK stars (dwarfs as well as giants) the stellar parameters derived with The Payne were consistent with isochrones and the published APOGEE DR14 values.

However, for the M~dwarfs the results were diverging from the isochrones and no comparison was done with APOGEE results, because APOGEE DR14 did not provide parameters for these stars.
Possible explanations given by \citet{The_Payne_Ting2019ApJ} for their mismatch compared with the isochrones are that the adopted line list was not well calibrated for this temperature range and that the used atmospheric models were not suitable for M~dwarfs. When testing how well The Payne recovered labels it was found that the deviation from the input labels was about twice as large for stars between 3000 and 4500~K than for hotter stars (see Fig.~5 in \citealt{The_Payne_Ting2019ApJ}). To improve the performance of The Payne framework for M~dwarfs, it is important to use a line list adapted for M~dwarfs, and the algorithm needs to be retrained on a set of synthetic spectra that better represent low-mass stars. 

\subsubsection{Model spectra and neural network}
\label{subsubsec:method:spectra}

In this study we followed a similar procedure for computing the model spectra and training the neural network as demonstrated in \citet{2019A&A...628A..54K}. 
The Payne's ANN was trained to restore the model spectrum corresponding to input labels.
The ANN architecture consists of a fully connected three-layer model with nine input units, two hidden layers with 400 and 300 units, and 11\,000 output units.
The number of input units corresponds to the dimensionality of the spectral grid used for training, while the number of output units corresponds to the number of wavelength points of the training spectra.
A ReLU (rectified linear unit) activation function is used for the hidden units, while a sigmoid activation function is used for the output units.

For training and validation a random uniform grid of synthetic spectra was computed using Turbospectrum as described in \citet{2022arXivGerber} together with MARCS atmospheric models, the APOGEE DR16 line list \citep{APOGEE_DR16_linelist_Smith2021}, and the water line list by \citet{2018MNRASPolyansky}.
The grid covers the stellar parameter space of $\teff$ from 2500 to 5500~K, $\lggu$ from 4 to 5.4~dex\footnote{The units of surface gravity are $\mathrm{cm\,s^{-2}}$. However, throughout the article, we use the unit dex when specifying values of $\lgg$.}, $\feh$ from $-2.0$ to 0.6~dex\footnote{Throughout the article the metallicity is represented by $\feh$. The abundances of the $\alpha$-elements are scaled as follows in the MARCS atmospheric models: $\afe = +0.4$ for $\feh \leq -1.0$, $\afe = -0.4 \times \feh$ for $-1.0 \leq \feh \leq 0.0$, and $\afe = 0.0$ for $\feh \geq 0.0$.}, and microturbulence $\vmic$ from 0.01 to 2.0~km s$^{-1}$.
In addition, the elemental abundances of O, Mg, Ca, Si, and Ti vary within $-0.2$ and 0.8~dex relative to the solar mixture \citep{2007SSRv..130..105G}, 
such that the distribution of each abundance ratio with respect to iron 
is uniform across the grid.
The model grid consists of 11\,292 spectra in total, of which 70\% were used for training and the remainder for validation.
For this study, the stellar spectra were modelled assuming local thermodynamic equilibrium (LTE). Apart from Gaussian instrumental broadening corresponding to the spectral resolution of APOGEE no additional broadening of spectral lines (e.g., $\vsini$) was applied.
We note that departures from LTE as well as rapid rotation can occur for M~dwarfs, as discussed in Sects.~\ref{sec:future:nlte} and \ref{sec:future:rotation}. However, we leave a possible non-LTE analysis and the fitting of $\vsini$ to future follow-up studies.

We performed a validation of the neural network by comparing synthetic spectra from the validation sample with models generated with the neural network using the same set of parameters. We found that the median interpolation error for the majority of models is about 0.1~\%. A few models deviate by more than 1~\%. The warmer parts of the grid (i.e., $\teff >$ 4000~K) perform slightly better than the cooler parts of the grid (below 0.1~\% compared with slightly above 0.1~\%). This can be seen in Fig.~\ref{Fig:histogram} in Appendix~\ref{append:validation}. The performance of the grid is similar to that reported by \citet{The_Payne_Ting2019ApJ}, who give a median interpolation error of about 0.1~\% for their grid and a slightly larger interpolation error for the cooler models.

The spectroscopic module uses normalised observed spectra (see Sect.~\ref{subsubsec:method:normalisation}) and a gradient descent method to find the global minimum in the parameter space of the training labels. In the process, the observed spectrum is compared to synthetic spectra reconstructed using the neural network, following the methodology in \cite{Gent_2022A&A}.

\subsubsection{Normalisation including pseudo-continuum}
\label{subsubsec:method:normalisation}

\begin{figure*}
   \centering
   \includegraphics[width=18cm]{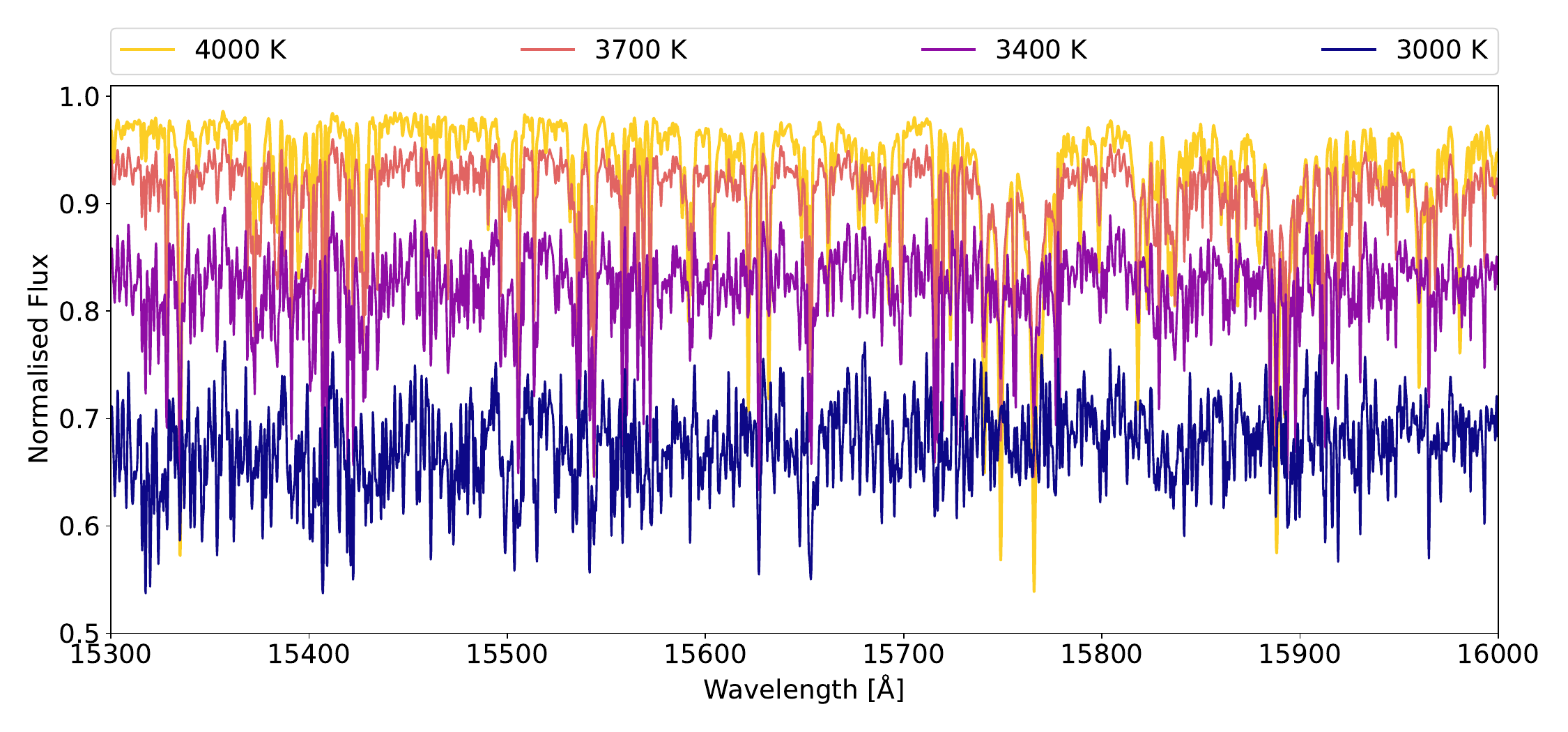}
   \caption{Example of H-band synthetic spectra generated with different effective temperatures. The surface gravity was set to 4.7~dex and the metallicity was set to solar.
   The different colours correspond to different $\teff$ values.}
   \label{Fig:Water_cont_Teff}
\end{figure*}

The synthetic spectra used for training the SAPP's ANN are normalised to the continuum flux. Thus, the SAPP needs normalised observed spectra in order to analyse the stars. However, M~dwarf spectra do not have a clear continuum due to the presence of a multitude of molecular lines, both in the optical and, to a lesser extent, in the NIR range. These molecular features suppress the continuum, forming a pseudo-continuum, where the suppression becomes deeper for cooler stars. In Fig.~\ref{Fig:Water_cont_Teff} we show how the pseudo-continuum varies with the effective temperature.
The synthetic spectra in the figure were generated in the same way as described in Sect.~\ref{subsubsec:method:spectra}.
In the wavelength region shown the highest flux points of the hottest synthetic spectra are almost 0.3 continuum units larger than the highest flux points of the coolest spectra. We also explored how the pseudo-continuum is affected by metallicity. We found that the flux depression is more severe at higher metallicity, although the effect remains within about 0.03 continuum units. This can be seen in Fig.~\ref{Fig:Water_cont_met_append} in Appendix~\ref{append:pseudocont}.
For a discussion of the effect of carbon and oxygen abundances on the pseudo-continuum, see \citet{2016ApJ...828...95V}. 

The goal is to match the scaling of the observed and synthetic spectra so that they can be compared.
The SAPP has a built-in normalisation routine which is described in the appendix of \citet{Gent_2022A&A}. The routine can be summarised as a piece-wise linear regression algorithm, whereby the un-normalised spectrum is broken into segments. These are defined considering the location of broad and narrow lines. Each segment is fitted taking into account the S/N, and the observed flux is divided by the fit in each segment. Due to the presence of the pseudo-continuum in M~dwarfs this normalisation procedure needed to be modified.
In order to apply a pseudo-continuum to the observed M~dwarf spectra that matches the pseudo-continuum level of the model grid spectra that were used in training The Payne (see Sect.~\ref{subsubsec:method:spectra}), we generated a grid of synthetic spectra in the same way as those shown in Fig.~\ref{Fig:Water_cont_Teff}
in the $\teff$ range of 3000 to 4150~K (it is limited to early- to mid-M~dwarfs) with a step size of 50~K, $\lgg$ of 4.4 to 4.9~dex, and metallicity of $-0.8$ to 0.5~dex\footnote{The range in metallicity covers the expected metallicities of the stars in the reference sample. In the synthetic spectra used for the development of the normalisation procedure, the abundances of the $\alpha$-elements are scaled as in the MARCS atmospheric models.}, where $\lgg$ and $\feh$ both had a step size of 0.1~dex. We included some variation in the surface gravity in the grid even though its main effect on the spectra is a broadening of the lines, with a minimal influence on the pseudo-continuum.
We fit second-degree polynomials to the highest peaks of the generated synthetic spectra disregarding peaks found outside of three sigma from the mean flux of the highest peaks.
This results in sets of polynomial coefficients for the different stellar parameters that are used as input to the SAPP. These polynomials are then used to adjust the flux of the observed spectra to the pseudo-continuum level.

The normalisation in the SAPP for M~dwarfs is done in parallel with parameter determination by first running the spectroscopic module on the observed spectra, where the observed spectra are treated the same way as for FGK stars, i.e. using the original SAPP normalisation routine. We use the best-fit $\teff$ and $\feh$ values together with the $\lgg$ obtained from the photometric module to find the corresponding polynomial derived from the grid of synthetic spectra. This polynomial is then multiplied with the original normalised observed spectrum which is then analysed again.
This process is repeated and iterated until convergence. We define convergence in the following way. We take the differences of parameters between each successive iteration. The differences decrease until they reach zero or the derived parameters oscillate between two fixed sets of values. This oscillation occurs for a minority of stars and was found to be stable for at least 100 iterations. By inspecting the stars in our sample, we found that $n$ = 10 was sufficient as an iteration limit, that is, the difference between steps reached zero or a constant value before $n$ iterations. Uncertainties are based on the values at the iteration step for which convergence is achieved (see Sect.~\ref{sec:method:uncertainties}), and the size of the oscillation, if any, contributes to the final spectroscopic uncertainty.

\subsubsection{Line mask}
\label{subsubsec:method:linemask}

\begin{figure*}
   \centering
   \hspace*{-2.5cm}
   \vspace*{-1cm}
   \includegraphics[width=23cm]{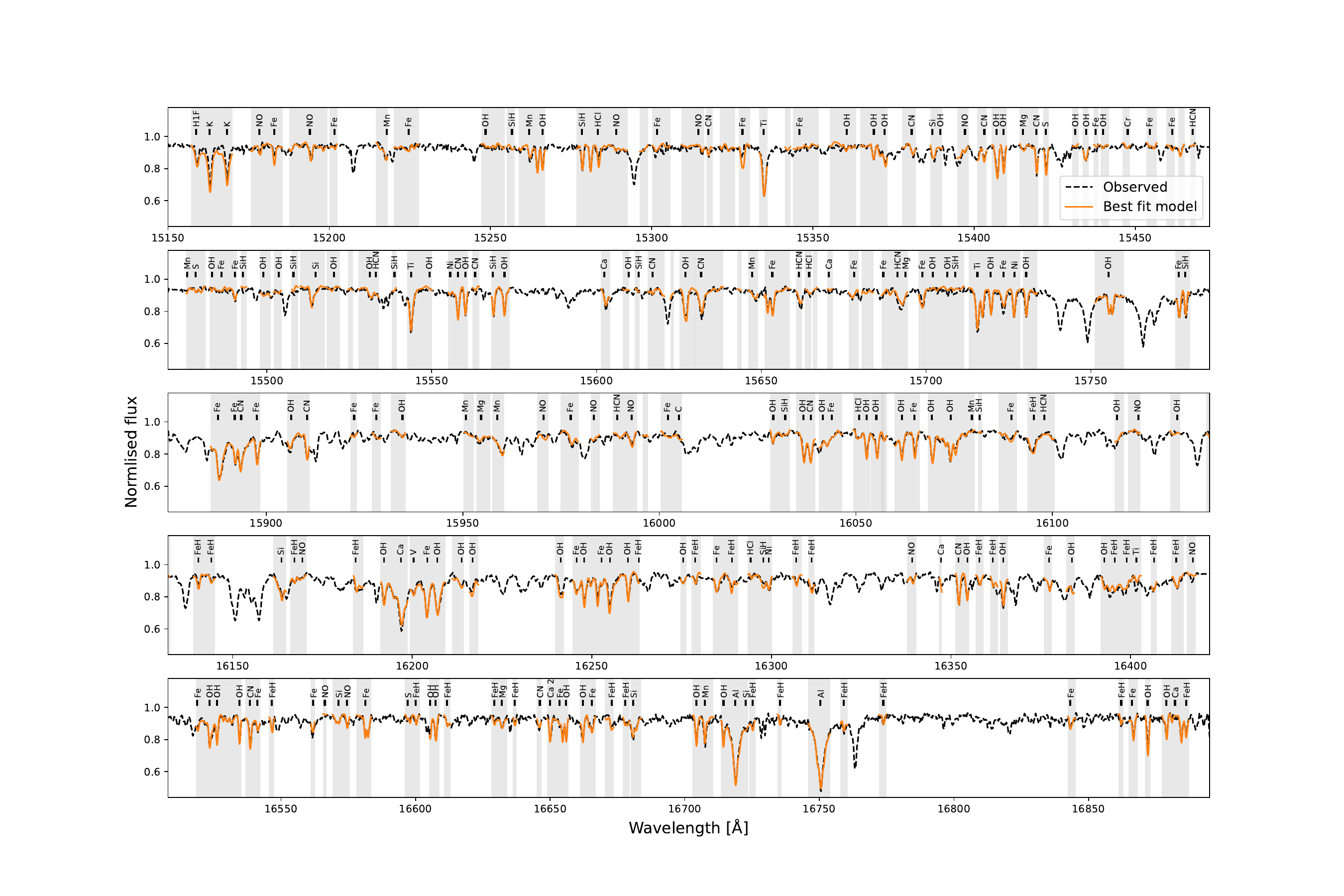}
   \caption{Normalised observed spectrum of the star GJ~880 as black dashed line, and best-fit model (synthetic spectrum predicted by the Payne's ANN for the parameters given in Table~\ref{tab:Result_analysis}) as orange solid line. Grey shaded areas indicate the location of the line mask we used. Derived parameters for this star are $\teff$:~3649~K, $\lgg$:~4.8~dex, and $\feh$:~0.25~dex.}
   \label{Fig:GJ880_spec}
\end{figure*}

First tests using the complete spectral range of the APOGEE data resulted in derived effective temperatures which were higher by more than 200~K compared to interferometric values for some stars.
In addition, a degeneracy between the effective temperature and metallicity was apparent.
Inspection of the fits indicated that this was due to some spectral regions which cannot be modelled well, which is compensated for by an inadequate change in stellar parameters.
To remedy this, we restricted the application of the fitting procedure to selected spectral ranges within a line mask. The line mask was taken from \citet{APOGEE_Sarmento2021}, who compared an observed spectrum of the M4V star Ross~128 (GJ~447) with a synthetic spectrum generated for parameters corresponding to this star. The construction of the line mask is described in their Sect.~3.3. For an illustration see Fig.~\ref{Fig:GJ880_spec}.


\subsection{SAPP version for M~dwarfs}
\label{sec:SAPP_lite}

\begin{figure}
   \includegraphics[width=0.51\textwidth,trim= 0 0 0 77,clip]{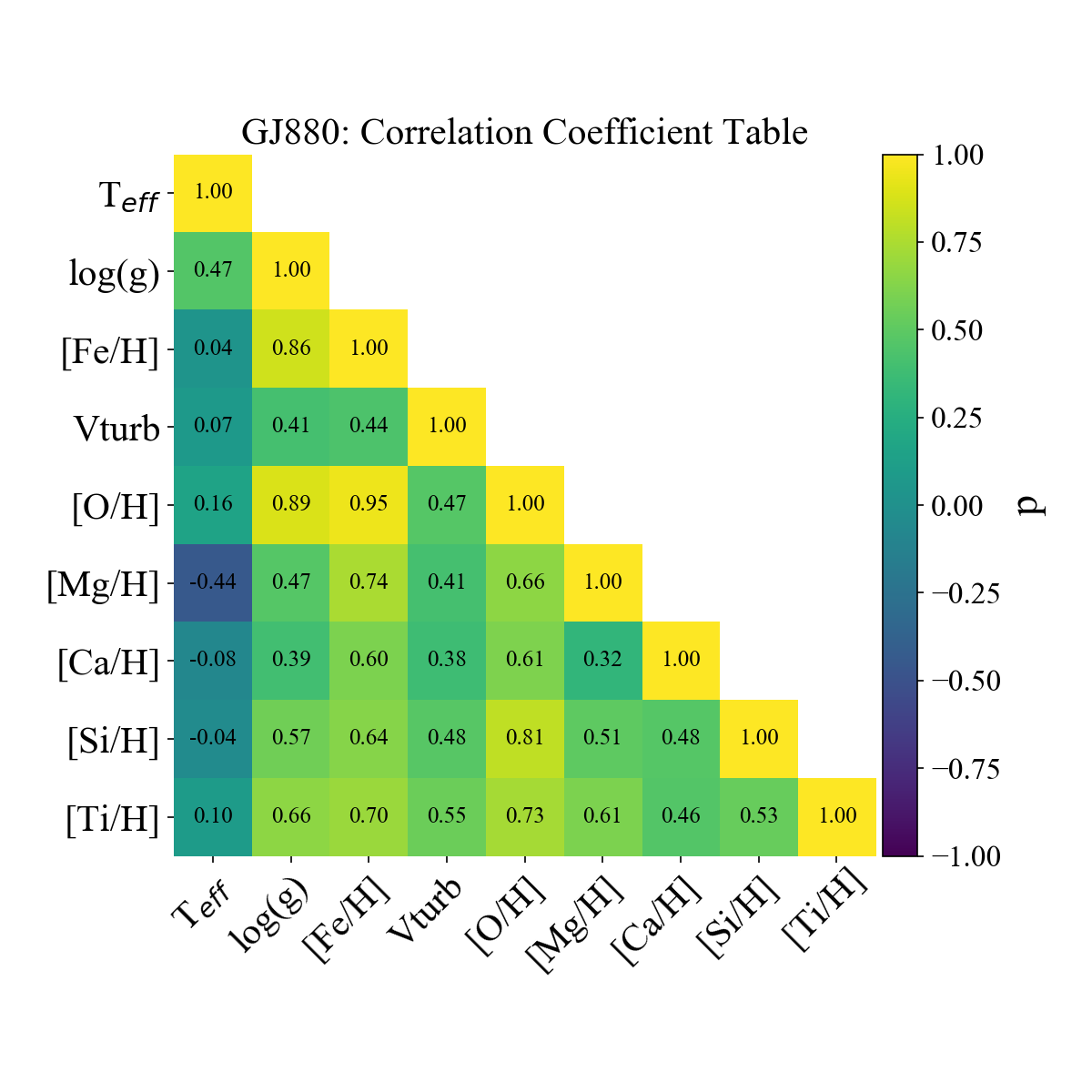}
   \caption{Correlation matrix for SAPP's spectroscopic module without photometric constraint for star GJ~880. The colour scale represents statistical correlation from $-1$ to 1 for nine ANN parameters.}
   \label{fig:spec_val_corr}
\end{figure}

The final parameters for M~dwarfs are derived via the spectroscopic module of the SAPP driven by the photometric surface gravity. Specifically, the most probable $\lgg$ value derived from the photometric PDF (see Sect.~\ref{subsec:method:photometry and evolution}) is passed on to the spectroscopic module and is fixed during the spectrum fitting process.
This approach helps to mitigate the strong degeneracies found between $\lgg$ and other parameters when applying the spectroscopic module alone. Figure~\ref{fig:spec_val_corr} shows the covariances for a free fit of the nine parameters in the spectroscopic module for GJ~880, a representative M~dwarf in our test sample. The correlations with surface gravity are among the most significant in the figure.
Thus, alternative constraints on the parameters are needed.
We note that since the ANN model includes a variation in individual element abundances, these are by default given in the output of the SAPP. However, we leave the validation of the derived abundances for M~dwarfs to future work.

\begin{figure*}
   \centering
   \hbox{\includegraphics[width=0.5\textwidth]{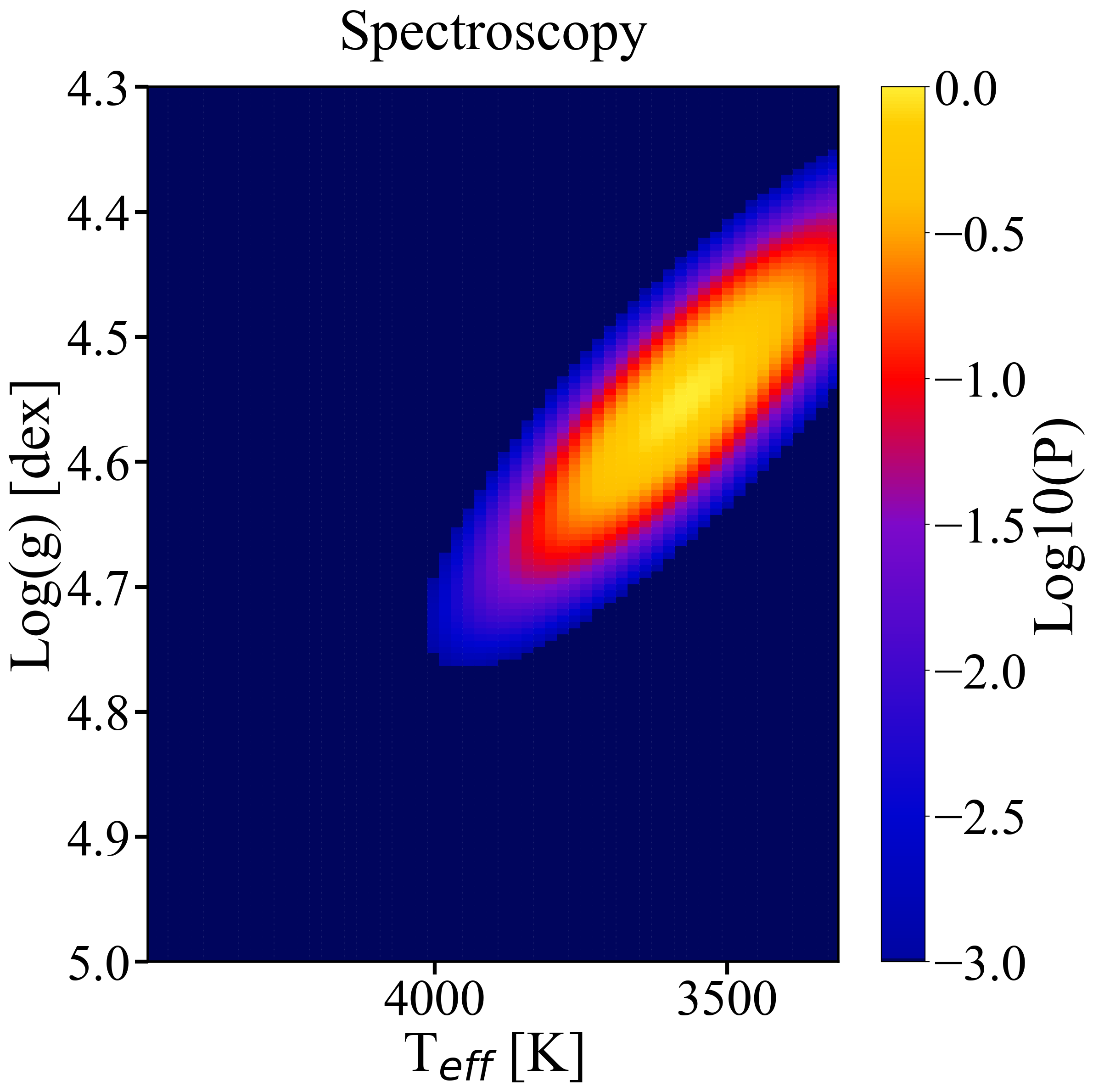}
\includegraphics[width=0.5\textwidth]{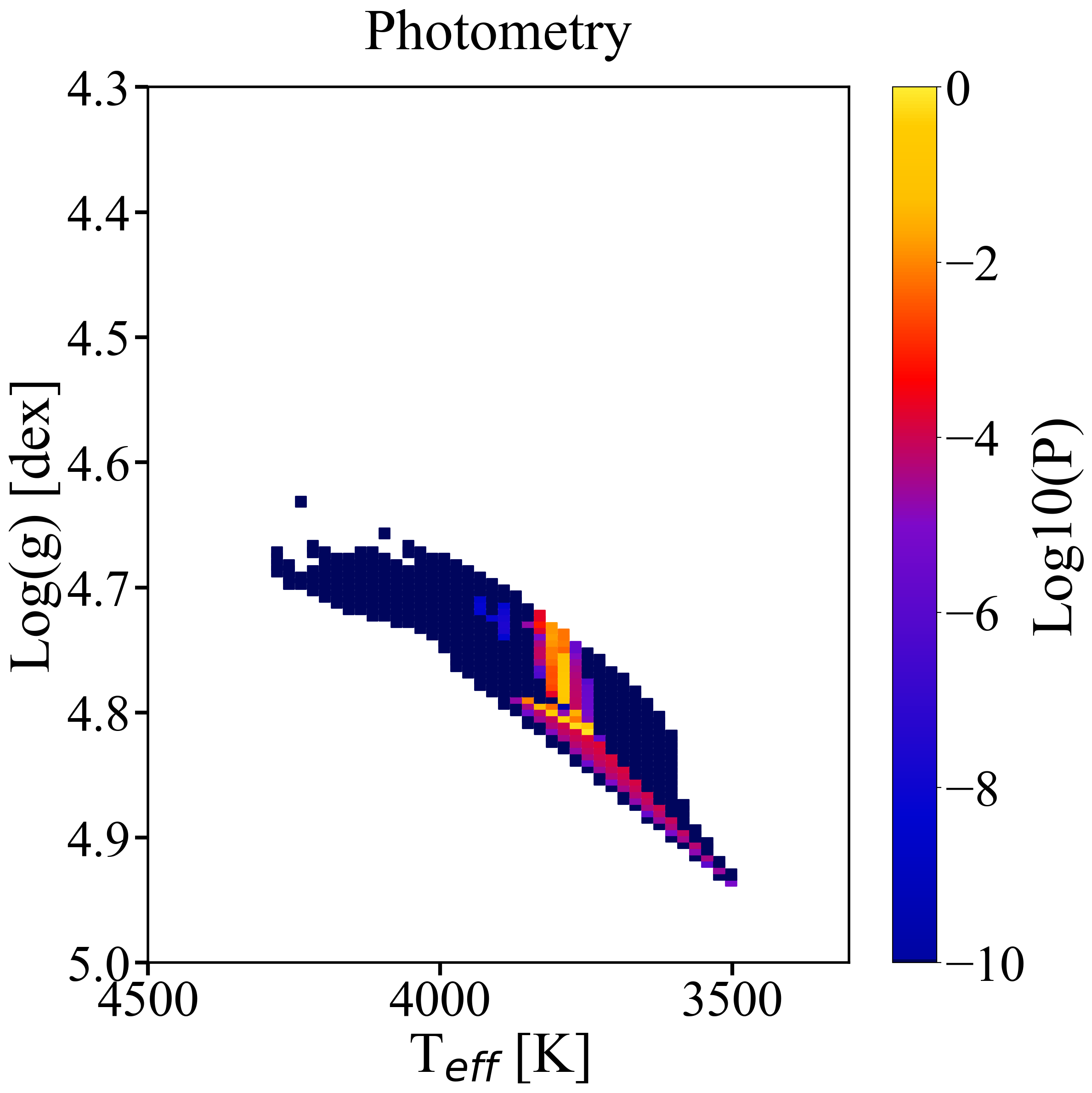}}
   \caption{PDFs calculated for GJ~880 for two different SAPP modules: spectroscopy (left) and photometry (right). The horizontal axis is effective temperature, the vertical axis is surface gravity, and the colour scale is the logarithm of probability. Each PDF is sliced in the [Fe/H] dimension at their maximum probability. White space corresponds to NaN values.}
   \label{fig:heat_map}
\end{figure*}

Figure~\ref{fig:heat_map} shows the PDFs for GJ~880 from the two modules of the SAPP. Each PDF shows the likelihood landscape in $\teff$-$\lgg$ space at the best-fitting $\feh$, with the colour scale representing the logarithm of the probability.
The correlation between $\teff$ and $\lgg$ in the spectroscopy module is apparent, while the valid values resulting from the photometric module are restricted to a smaller fraction of the parameter space. The main visual differences in the probability space between the two can be accounted for by how they were calculated. The spectroscopy PDF is built from its best-fit $\teff$, $\lgg$, $\feh$, and the correlation matrix derived from curvefit (see Fig.~\ref{fig:spec_val_corr}). These parameters are then compared to the common atmospheric parameters shared by photometry ($\teff$, $\lgg$, and $\feh$) to build a PDF space. However, the photometry PDF is the $\teff$-$\lgg$ plane of a multi-dimensional set of isochrones. As in \citet{Gent_2022A&A}, the photometry PDF is similar in structure to an evolution track.
The value of the surface gravity is calculated as the average over all values from the photometric module within the subdomain defined in Sect.~\ref{subsec:method:photometry and evolution}, weighted by the corresponding probabilities.

\subsection{Calculated internal uncertainties}
\label{sec:method:uncertainties}
The SAPP includes a calculation of the internal uncertainties arising from the application of the algorithm.
These are shown together with our results in Table~\ref{tab:Result_analysis}.
For an estimation of the overall uncertainties based on comparison with external data from the literature see Sect.~\ref{subsec:result:uncertainties}.
The internal uncertainty in $\lgg$ is derived from the photometric PDF which propagates the apparent magnitude, distance and reddening uncertainties.
The $\lgg$ produced from the photometric module directly comes from the weighted average of the PDF. From this posterior distribution we derive a weighted standard deviation which is the reported uncertainty for this parameter.

The internal uncertainties in $\teff$ and $\feh$ are derived by propagating uncertainties from different parts of the SAPP quadratically as follows:
\begin{equation}
    u = \sqrt{ {(\sigma_{spec})}^2 + {(\sigma_{pseudo})}^2 +
    {(\sigma_{phot})}^2},
\end{equation}
where $\sigma_{spec}$ is the spectroscopic uncertainty for each parameter derived via the leading diagonal of the co-variance matrix\footnote{This is derived using the Python module \texttt{scipy.optimize.curve\_fit} \citep{2020SciPy-NMeth}.}, corresponding to the square root of the variance.
$\sigma_{pseudo}$ is the uncertainty arising from the iterative  normalisation procedure allowing for the pseudo-continuum, as described in Sect.~\ref{subsubsec:method:normalisation}.
More specifically, it is the standard deviation of the set of parameter values derived in each iteration until convergence.
%
%
$\sigma_{phot}$ is the uncertainty derived from propagating the uncertainty of the photometric $\lgg$ through the spectroscopic method. As the $\lgg$ used in the spectroscopic module is fixed to the maximum likelihood $\lgg$ from the photometric module, its uncertainty directly contributes to the uncertainties in $\teff$ and $\feh$.
Furthermore, we do not allow the fitting procedure to go beyond the bounds defined by the photometric $\lgg$ and its uncertainty.

We note that the procedure for estimating the uncertainties in the SAPP for M~dwarfs is different from that used in the SAPP for FGK stars, as the full Bayesian inference scheme is not yet implemented. Furthermore, contrary to what is described in Sect.~3.5.3 in \citet{Gent_2022A&A}, we do not apply an ``error model'', owing to the limited number of reference stars.



\section{Results and discussion}
\label{sec:Results}


In this section we describe the results from applying the SAPP for M~dwarfs to our test sample in order to assess the performance of our pipeline.
We list our derived stellar parameters in Table~\ref{tab:Result_analysis}. The star LSPM~J1204+1728S is not included in this table as the derived parameters\footnote{The derived parameters for LSPM~J1204+1728S are $\teff$ = 3185 $\pm$171~K,  $\lgg$ = 4.859 $\pm$0.008~dex, $\feh$ = $-0.21$ $\pm$0.06~dex.} were judged to be unreliable due to its fast rotation (see below). We note that the star GJ~105A is an early K~dwarf, and our derived stellar atmospheric parameters might therefore be unreliable, as the pipeline is optimised for M~dwarfs. The star GJ~338A is classified as a late K~dwarf, which is closer to the parameter range targeted by the pipeline.
We note that, as mentioned in Sect.~\ref{subsubsec:method:spectra} and shown in Fig.~\ref{fig:spec_val_corr} the ANN model includes a variation the microturbulence parameter $\vmic$, and best-fit values are given in the output of the SAPP. However, in the context of this work we regard the microturbulence as a free nuisance parameter, as its physical meaning is limited and an evaluation with independent reference values is not possible. Summarising the fitting results for the sample as a whole, the $\vmic$ values show a rather flat distribution of values ranging from 0.03 to 1.55~$\kms$, with a median of 0.78~$\kms$.

In Fig.~\ref{Fig:GJ880_spec} we show an example of a best fit model for the star GJ~880 ($\teff$, $\log g$, [Fe/H] = 3650~K, 4.8~dex, 0.25~dex) in comparison with the normalised observed spectrum. Examples of best fit models for two additional stars can be found in Appendix~\ref{append:fits}. The line mask that is used in the SAPP for M~dwarfs is indicated in grey. The fit is generally good in the regions covered by the line mask. 
Lines found at the edges of the detectors have a slightly worse fit (the edges of the detectors are outside of the ranges shown in Fig.~\ref{Fig:GJ880_spec}). This worse fit is most likely caused by the normalisation routine which behaves worse at the edges of the detectors.
We note that APOGEE spectra suffer from persistence effects\footnote{Elevated counts with an amplitude related to the previous exposure.}, in particular at the shortest wavelengths \citep{2018AJ....156..125H}. 
We can also see that the two potassium lines at 15163~\AA\ and 15168~\AA\  show a slightly worse fit. These lines were shown to be affected by non-LTE effects in \citet{Olander2021} which could explain part of the mismatch.

\begin{table*}[t]
    \centering
    \caption{Stellar parameters and their uncertainties derived in this work.}
    \label{tab:Result_analysis}
    \begin{tabular}{lrrrrrr}
    \noalign{\smallskip}
    \hline
    \noalign{\smallskip}
    Star &  $\teff$/K &  $u_{\teff}$ &  $\lggu$ & $u_{\log g}$ &  [Fe/H] &  $u_{\rm[Fe/H]}$ \\
    \noalign{\smallskip}
    \hline
    \noalign{\smallskip} 
      BD+00549B &  3726 &      60 & 4.931 &     0.029 &  -0.66 & 0.07 \\
     BD-064756B &  3442 &      33 & 4.814 &     0.006 &  -0.02 & 0.06 \\
         GJ105A &  4607 &      93 & 4.960 &     0.150 &   0.02 & 0.06 \\
         GJ105B &  3483 &      69 & 4.994 &     0.017 &  -0.31 & 0.05 \\
          GJ15A &  3609 &      73 & 4.883 &     0.023 &  -0.30 & 0.16 \\
          GJ205 &  3941 &      26 & 4.727 &     0.008 &   0.19 & 0.06 \\
          GJ212 &  3838 &      35 & 4.734 &     0.015 &   0.09 & 0.10 \\
       GJ297.2B &  3597 &      52 & 4.801 &     0.004 &  -0.08 & 0.08 \\
         GJ3195 &  3571 &      52 & 4.863 &     0.014 &  -0.10 & 0.11 \\
         GJ324B &  3262 &      76 & 5.000 &     0.003 &   0.24 & 0.11 \\
         GJ338A &  4000 &      19 & 4.690 &     0.012 &  -0.11 & 0.05 \\
         GJ338B &  4029 &      19 & 4.711 &     0.013 &  -0.13 & 0.05 \\
          GJ393 &  3586 &      47 & 4.847 &     0.018 &  -0.15 & 0.06 \\
         GJ412A &  3593 &      96 & 4.900 &     0.020 &  -0.31 & 0.12 \\
          GJ447 &  3243 &     101 & 5.066 &     0.015 &  -0.13 & 0.05 \\
          GJ526 &  3729 &      38 & 4.792 &     0.018 &  -0.32 & 0.08 \\
          GJ687 &  3539 &      69 & 4.875 &     0.063 &  -0.02 & 0.12 \\
         GJ725A &  3584 &      51 & 4.925 &     0.025 &  -0.34 & 0.05 \\
         GJ725B &  3556 &      61 & 4.984 &     0.032 &  -0.38 & 0.06 \\
         GJ752A &  3684 &      38 & 4.819 &     0.035 &  -0.09 & 0.08 \\
         GJ777B &  3251 &     152 & 5.043 &     0.038 &  -0.08 & 0.04 \\
          GJ809 &  3839 &      41 & 4.741 &     0.054 &  -0.12 & 0.11 \\
          GJ880 &  3649 &      83 & 4.754 &     0.009 &   0.25 & 0.12 \\
 LSPMJ0355+5214 &  3481 &      61 & 4.986 &     0.014 &  -0.19 & 0.02 \\
        Ross799 &  3686 &      29 & 4.790 &     0.007 &  -0.04 & 0.05 \\
   \noalign{\smallskip}
   \hline
   \end{tabular}
   \tablefoot{The M~dwarf LSPM~J1204+1728S is not included (see text). GJ~105A and GJ~338A are K~dwarfs but are included since they are within the parameter range of the model grid used to train The Payne and have APOGEE spectra. The uncertainties are derived as described in Sect.~\ref{sec:method:uncertainties}.}
\end{table*}

\begin{figure}
   \centering
   \includegraphics[width=8cm]{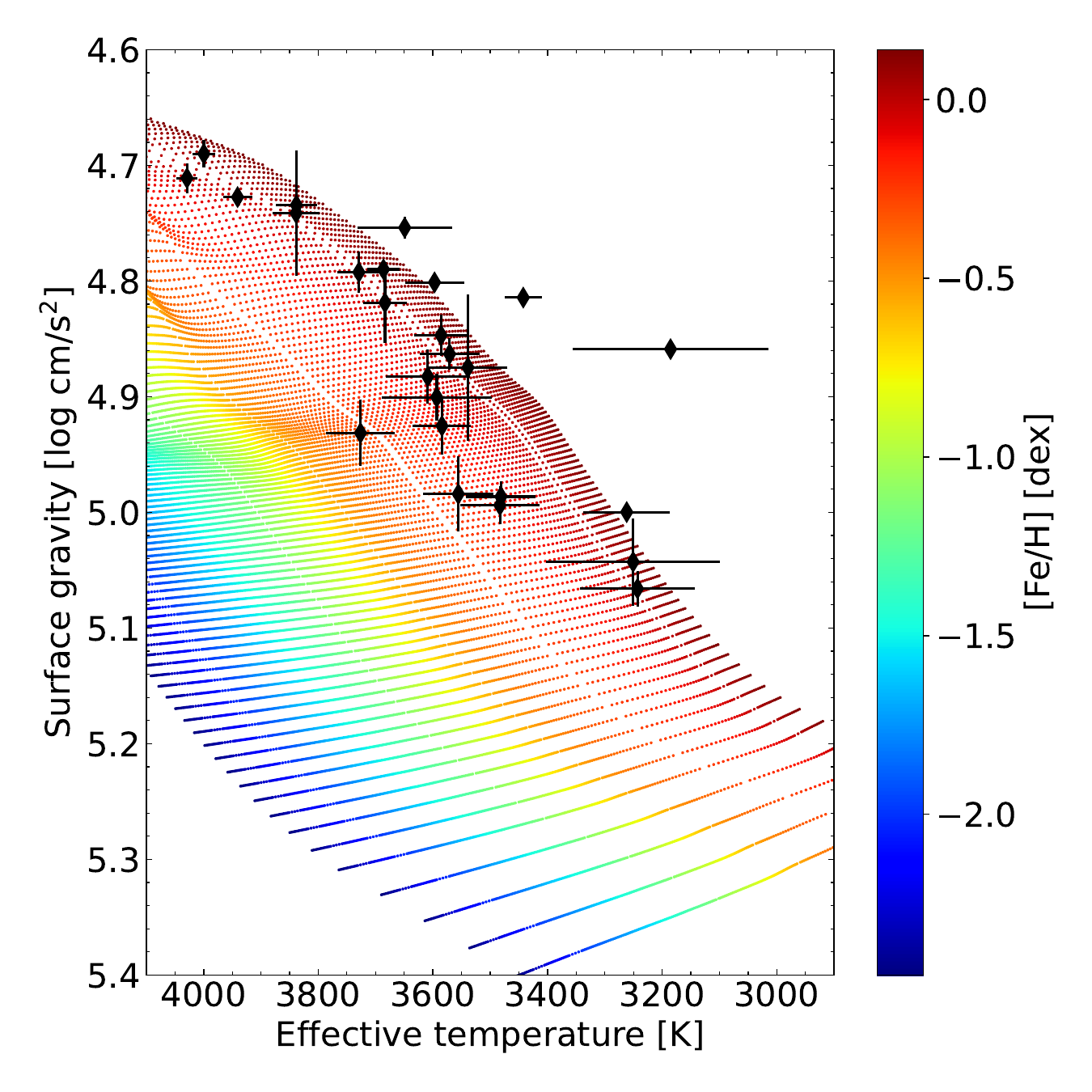}
   \caption{Surface gravity versus effective temperature derived by the SAPP (black diamonds with error bars). The K~dwarf GJ~105A is not visible since its parameters are outside of the axis ranges. Small dots represent a subset of the grid of stellar evolution models used by the photometric module as described in Sect.~\ref{subsec:method:photometry and evolution}, selected for this illustration to have an age of 13~Gyr, colour-coded by metallicity, with masses increasing from 0.1~M$_\odot$ at the lower right towards the upper left with steps of 0.005~M$_\odot$. We note that when constructing the PDF for surface gravity the photometric module uses the whole grid of models for all available ages.}
   \label{Fig:HRD}
\end{figure}

In Fig.~\ref{Fig:HRD} we show the $\teff$ values derived from spectroscopy and the $\log g$ values derived from photometry together with the parameters covered by a subset of our grid of stellar evolution models (Sect.~\ref{subsec:method:photometry and evolution}). The models shown correspond to an age of 13~Gyr (the maximum age that is physical for a star) and are colour-coded by metallicity. 
We recall that when constructing the PDF for surface gravity the photometric module uses the whole grid of models for all available ages.
We also note that the stars lie in the region of models colour-coded in red, in agreement with our derived metallicities of $-0.7$~dex and higher.
The star lying furthest outside of the parameter space covered by the evolutionary model grid is the fast rotator LSPM~J1204+1728S, discussed in Sect.~\ref{subsec:result:comparison:spectroscopy}.
The second outlier in the same sense is BD$-$06~4756B, indicating that the uncertainty of the effective temperature and/or surface gravity might be underestimated for this star.

In the sections below we compare our results with literature values from some of the works mentioned in Sect.~\ref{sec:introduction} based on several different techniques. The aim is to cross-check our method, and to understand its accuracy, precision, and scope of applicability.


\subsection{Comparison based on interferometry}
\label{subsec:result:comparison:interferometry}

We used \citet{Boyajian2012} and \citet{Rabus2019} to obtain reference parameters based on interferometric measurements.
\citet{Boyajian2012} used the CHARA array to obtain limb-darkened angular diameters \diam. Coupled with Hipparcos parallaxes \citep{2007A&A...474..653V} and photometry fitted to spectral templates in order to obtain the bolometric fluxes, they calculated stellar radii and effective temperatures. They obtained stellar masses using an absolute K-band mass-luminosity relation from \citet{1993AJ....106..773H}. 
\citet{Rabus2019} used the VLTI/PIONIER interferometer to obtain \diam. They used \gaia\ DR2 parallaxes in their analysis, and bolometric fluxes were obtained by integration over stellar model spectra fitted to photometric observations. For the masses they used an empirical mass-luminosity relation from \citet{2019ApJ...871...63M}. 

In our sample, 12 stars have angular diameter measurements by \citet{Boyajian2012} and one by \citet[][GJ~447]{Rabus2019}. In addition, \citet{Rabus2019} calculated effective temperatures and radii for five stars using uniform-disk angular diameters from \citet{Boyajian2012} and using their own determinations as described above otherwise (indicated in Table~\ref{tab:Sample} as ``B, R'' in column ``Int.'')\footnote{Due to using different limb darkening coefficients the values for \diam\ by \citet{Rabus2019} are lower than those of \citet{Boyajian2012} by about 1\% for these stars (2\% for GJ~205). Similarly, the absolute difference in bolometric flux ranges from about 1 to 4\%.}.
We calculated the surface gravities using the masses $M$ and radii $R$ given in \citet{Boyajian2012}\footnote{Using \gaia\ parallaxes and \citet{Boyajian2012} angular diameters results in stellar radii in agreement with the published values at the 1\% level (except for the binary GJ~338).} and \citet{Rabus2019}, and the equation $g = GM/R^2$.

\begin{figure} 
   \centering
   \includegraphics[width=8cm]{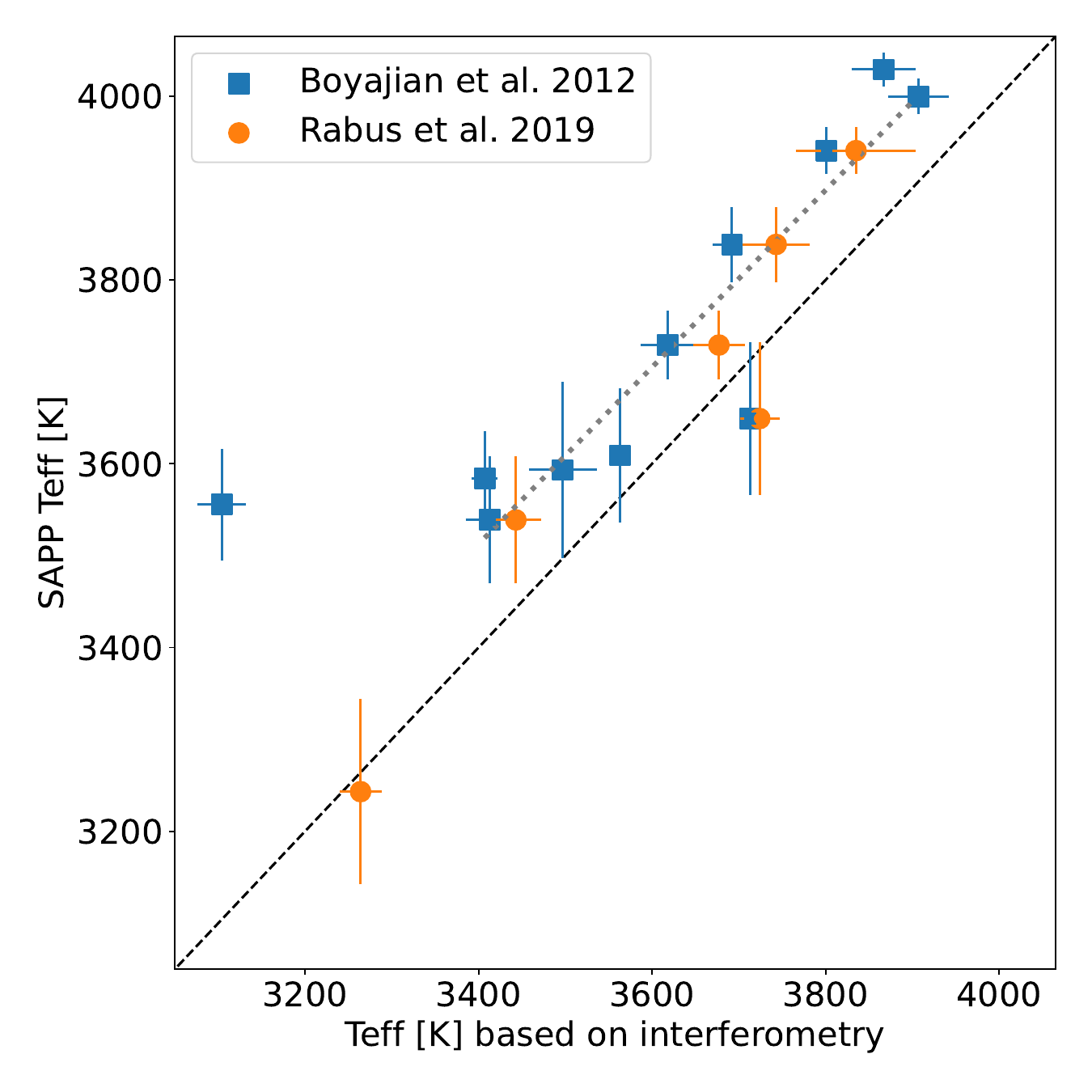}
   \includegraphics[width=8cm]{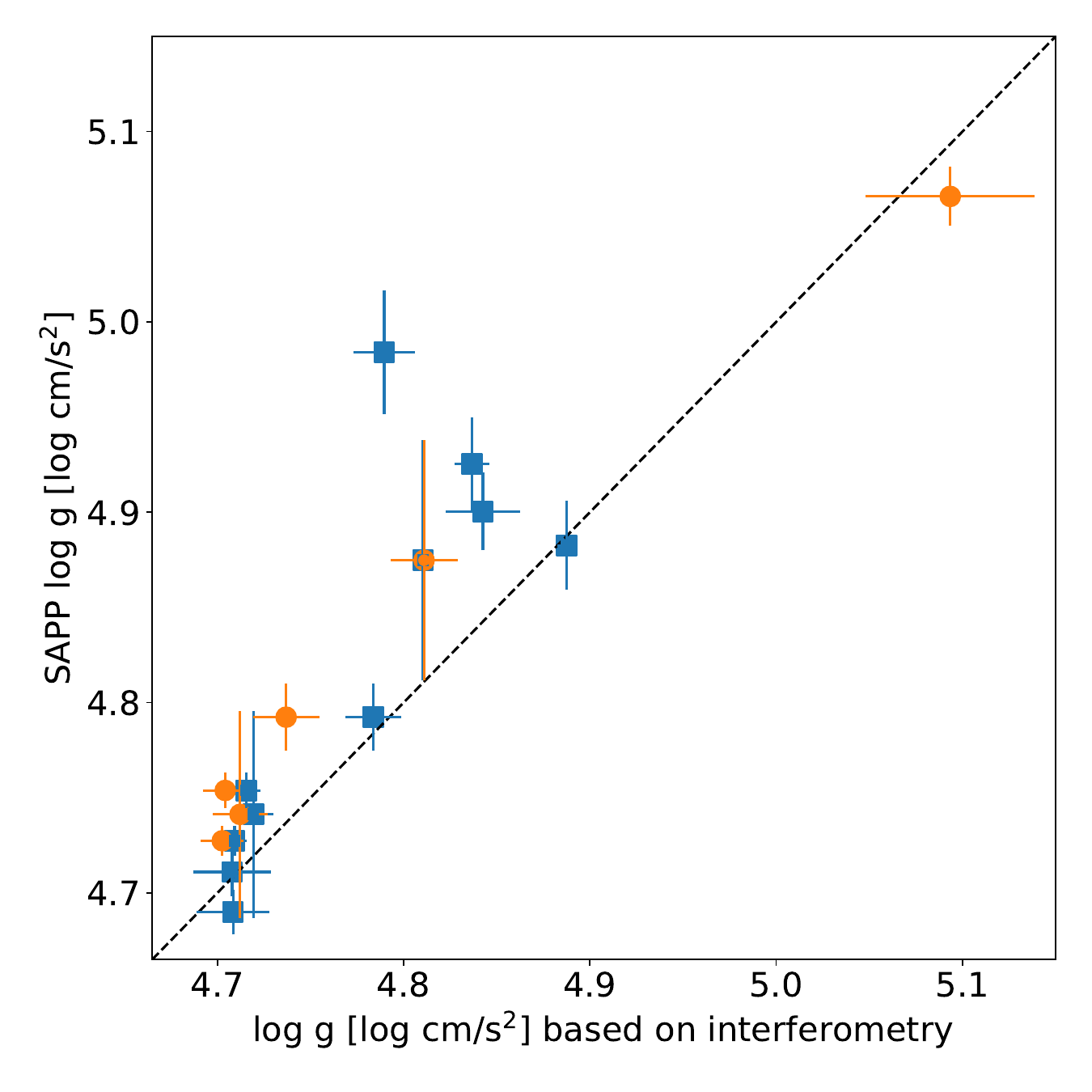}
   \caption{Comparing $\teff$ (top) and $\lgg$ (bottom) derived from the SAPP with corresponding parameters based on interferometric angular diameters \citep{Boyajian2012,Rabus2019}. The black dashed line in both figures corresponds to the 1:1 ratio and the grey dotted line in the top panel corresponds to a linear fit to the values from \citet{Boyajian2012}. We excluded the outlier GJ~725B for which the SAPP derived a $\teff$ of roughly 3550~K (leftmost blue square in top panel) from the linear fit, for reasons discussed in the text. We note that the K3 dwarf GJ~105A with $\teff\sim$4600~K is not shown in the figures.} 
   \label{Fig:Interferometry}
\end{figure}

In Fig.~\ref{Fig:Interferometry} we compare the $\teff$ and $\lgg$ values derived with the SAPP with the parameters based on interferometric angular diameters.
As mentioned above, \citet{Rabus2019} re-calculated $\teff$ using their method and angular diameters from \citet{Boyajian2012} for five of the stars. 
Therefore the two sets of results are not completely independent.
The effective temperature shows a clear linear trend with both \citet{Boyajian2012} and \citet{Rabus2019}. Our derived $\teff$ is generally higher in comparison with \citet{Boyajian2012}. When comparing with \citet{Boyajian2012} we find a strong outlier -- the star GJ~725B, for which \citet{Boyajian2012} derived 3104~K and we obtained 3556~K.
Other studies have also derived a higher $\teff$ for this star than what was obtained by \citet{Boyajian2012}. \citet{APOGEE_Sarmento2021} derived a temperature of 3544~K, \citet{2020AAMaldonado} 3291~K, \citet{2015ApJMann} 3345~K, and \citet{2022ApJSouto} 3400~K.
In addition, in most of the studies mentioned above the derived $\teff$ values for the two binary components GJ~725A and GJ~725B are the same to within 100~K, including our results (difference of 28~K). However, \citet{Boyajian2012} derived a difference of about 300~K in $\teff$ between the two stars. It is clear that GJ~725B needs further investigation which is beyond the scope of this work.

We performed a linear fit to the data from \citet{Boyajian2012}, excluding GJ~725B for the reasons mentioned above. The linear fit (with slope 0.963 and intercept 238~K) shows an offset of approximately 100~K above the 1:1 line (see Fig.~\ref{Fig:Interferometry}).
The mean absolute difference (MAD)\footnote{Here defined as MAD = $\sum_{i=1}^N |P_i({\rm SAPP})-P_i({\rm lit})|/N$, where $P$ is the stellar parameter compared, and $N$ is the number of comparison values.}
in effective temperature is 116~K. 
As can be seen in the figure, \citet{Rabus2019} derived slightly higher $\teff$ values for the five stars for which they re-analysed the data from \citet{Boyajian2012}, although they agree within uncertainties.
When comparing with \citet{Rabus2019} we find a MAD of 74~K. 
For the majority of the stars in the sample we derived a higher $\teff$ than both interferometric studies.
On the other hand, we do not see a systematic difference with other spectroscopic studies, as shown in Sect.~\ref{subsec:result:comparison:spectroscopy} and Fig.~\ref{Fig:Spec_param}. This implies that the offset seems to be a general trend when comparing spectroscopy and interferometric measurements.
Therefore, it could be due to the modelling components used in interferometry, such as accounting for limb-darkening or methods for measuring the bolometric flux. An indication of this can be seen in Fig.~\ref{Fig:Interferometry} as the recalculated effective temperatures by \citet{Rabus2019} are higher than those of \citet{Boyajian2012}. In this case, the change in modelling has decreased the offset between spectroscopic and interferometric $\teff$ values. A similar discussion can be found in \citet[][their Sect.~4.2 and Fig.~4]{2020ApJSouto}.

In the bottom panel of Fig.~\ref{Fig:Interferometry} we show the comparison between our derived surface gravity and the one calculated using the mass and radius from \citet{Boyajian2012} and \citet{Rabus2019}. We stress that the surface gravity based on the interferometric radius is not as fundamental as the effective temperature because empirical relations are used to calculate the mass of the stars. The stars in our sample generally follow the 1:1 ratio, with a possible small positive offset. The two stars in the binary mentioned above, GJ~725A and B, show the largest deviations of 0.1 (A) to 0.2 (B) dex.
The calculated MAD between $\lgg$ derived with the SAPP and by \citet{Boyajian2012} is 0.032 dex (excluding the outlier GJ~725B). 
The MAD for the surface gravity when comparing with \citet{Rabus2019} is 0.042 dex.
More interferometric measurements and direct mass determinations are needed in order to draw conclusions regarding the accuracy of the surface gravity derived by the SAPP.

We note that in a companion paper to \citet{Rabus2019}, \citet{2019MNRAS.484.2656L} 
used a method to estimate the uncertainties of the measured angular diameters that takes into account correlations between observables and includes systematic errors. This results in larger uncertainties than obtained by the standard method used for instance by \citet{Boyajian2012}, in particular for stars with small angular diameters ($\lesssim$0.6~mas, see Fig.~3 of \citealt{2019MNRAS.484.2656L}). However, for the stars in the samples of \citet{Boyajian2012} and \citet{2019MNRAS.484.2656L} which show the largest overlap in angular size ($\sim$0.7 to 0.8~mas) the uncertainties derived in the two works are comparable.


\subsection{Comparison with classical spectroscopy}
\label{subsec:result:comparison:spectroscopy}

\subsubsection{Reference values}
We compare our results with several spectroscopic studies of M~dwarfs, both in the optical and the NIR.
Starting with classical spectrum-fitting methods, we include the results of \citet{2015ApJMann} based on low-resolution spectra calibrated to an absolute flux scale in the optical and in the NIR, complemented by photometry and trigonometric parallaxes. These authors derived bolometric fluxes, $\teff$, metallicity, stellar radii, and stellar masses (using the empirical mass-luminosity relation from \citealt{2000A&A...364..217D}) 
for about 180 nearby K7 to M7 stars, including the majority of the stars in our sample. We calculated the surface gravity and corresponding uncertainty from the mass and radius given by \citet{2015ApJMann}.
Another reference study in the optical is \citet{2015A&A...577A.132M,2020AAMaldonado}, who applied an equivalent width analysis to HARPS and HARPS-N spectra \citep{2003Msngr.114...20M,2012SPIE.8446E..1VC} 
to obtain stellar parameters for about 200 M~dwarfs.
They used 13 and 47 stars of their sample for $\teff$ and metallicity calibration, respectively.

Turning to high-resolution spectroscopy, \citet{Pass2018,Pass2019} used spectra from the CARMENES spectrograph \citep{2014SPIEQuirrenbach_CARMENES} in both the optical and the NIR together with PHOENIX model atmospheres \citep{2013Husser_PHOENIX,Meyer17} to determine parameters for about 300 M~dwarfs.
\citet{Pass2019} give three sets of parameters based on spectra obtained in the optical, NIR, and both combined. We compare with the parameters derived using the combined spectra, judged by the authors to give the best results.
\citet{2022MNRAS.516.3802C} 
analysed spectra of 44 M~dwarfs obtained with the SPIRou spectrograph \citep{2020MNRAS.498.5684D}, using MARCS model atmospheres and Turbospectrum \citep{Turbospectrum_Alvarez1998}. They calibrated their method using 12 stars of their sample. Among the remaining stars there are five in common with our work, which we include in the comparison.
We also compare with two studies using the same instrument as our work.
\citet{APOGEE_Sarmento2021} used APOGEE spectra together with MARCS, Turbospectrum, and a custom line list built partly from the APOGEE line list \citep{2015ApJShetron_APOGEElinelist} to derive parameters for about 300 M~dwarfs.
\citet[][and references therein]{2022ApJSouto} also used APOGEE spectra and the behaviour of oxygen abundances as a function of $\teff$ and $\lgg$ to obtain those parameters for a sample of 21 stars.

A summary of the studies mentioned above can be seen in Table~\ref{tab:referenceStudies}. The table lists the references for the published parameters, the instruments used to obtain the spectra, the wavelength ranges, the model atmosphere or technique adopted, and the parameters obtained.
For more details regarding the analyses the reader is directed to the individual publications listed in the table.

\begin{table*}[ht]
   \caption{Spectroscopic reference studies. Top part: classical spectroscopy, bottom part: machine-learning approaches.}
   \label{tab:referenceStudies}
   \centering
   \begin{tabular}{l l l l l c}
        \noalign{\smallskip}
        \hline
        \noalign{\smallskip}
        Reference & Instrument & Wavelength & Resolving & Model/ \\
                  &            & [\AA]      & power $R$ & Method \\
        \noalign{\smallskip}
        \hline
        \noalign{\smallskip}
        \citet{2022MNRAS.516.3802C} & SPIRou & 9670-23200 & 70\,000 & MARCS \\
        \citet{2015ApJMann} & SNIFS & 3200-9700 & 1000 & SED fitting \\
        \citet{2015ApJMann} & SpeX & 8000-24000 & 2000 & SED fitting \\
        \citet{2020AAMaldonado} & HARPS/HARPS-N & 5300-6900 & 115\,000 & Pseudo EW\tablefootmark{d} \\
        \citet{Pass2019}\tablefootmark{a} & CARMENES & 7000-15200 & $>80\,000$\tablefootmark{c} & PHOENIX \\
        \citet{APOGEE_Sarmento2021} & APOGEE & 15000-17000 & 22\,500 & MARCS \\
        \citet{2022ApJSouto} & APOGEE & 15000-17000 & 22\,500 & MARCS \\
        \noalign{\smallskip}
        \hline
        \noalign{\smallskip}
        \citet{The_Cannon_Mdwarf_Birky2020ApJ}\tablefootmark{b} & APOGEE & 15000-17000 & 22\,500 & The Cannon \\
        \citet{2022AaA...658A.194P} & CARMENES & 8800-8835 & $>80\,000$\tablefootmark{c} & Deep Learning A \\
        \citet{2022AaA...658A.194P} & CARMENES & 6477-12816 & $>80\,000$\tablefootmark{c} & Deep Learning C2 \\
        \noalign{\smallskip}
        \hline
   \end{tabular}
   \tablefoot{
   \tablefoottext{a}{In \citet{Pass2019} the $\lgg$ is given by evolutionary models (PARSEC) corresponding to $\teff$ and $\feh$ at each step of the spectrum fit.}
   \tablefoottext{b}{\citet{The_Cannon_Mdwarf_Birky2020ApJ} only derived the parameters $\teff$ and $\feh$, all other studies $\teff$, $\lgg$, and $\feh$.}
   \tablefoottext{c}{The resolving power for CARMENES is $R \sim 94\,600$ and $R \sim 80\,500$ in the visible and NIR, respectively.}
   \tablefoottext{d}{\citet{2020AAMaldonado} used ratios of pseudo-equivalent widths of spectral features described in \citet{2015A&A...577A.132M}.}
   }
\end{table*}

\subsubsection{Effective temperature}

\begin{figure*}
   \centering
   \includegraphics[width=6cm]{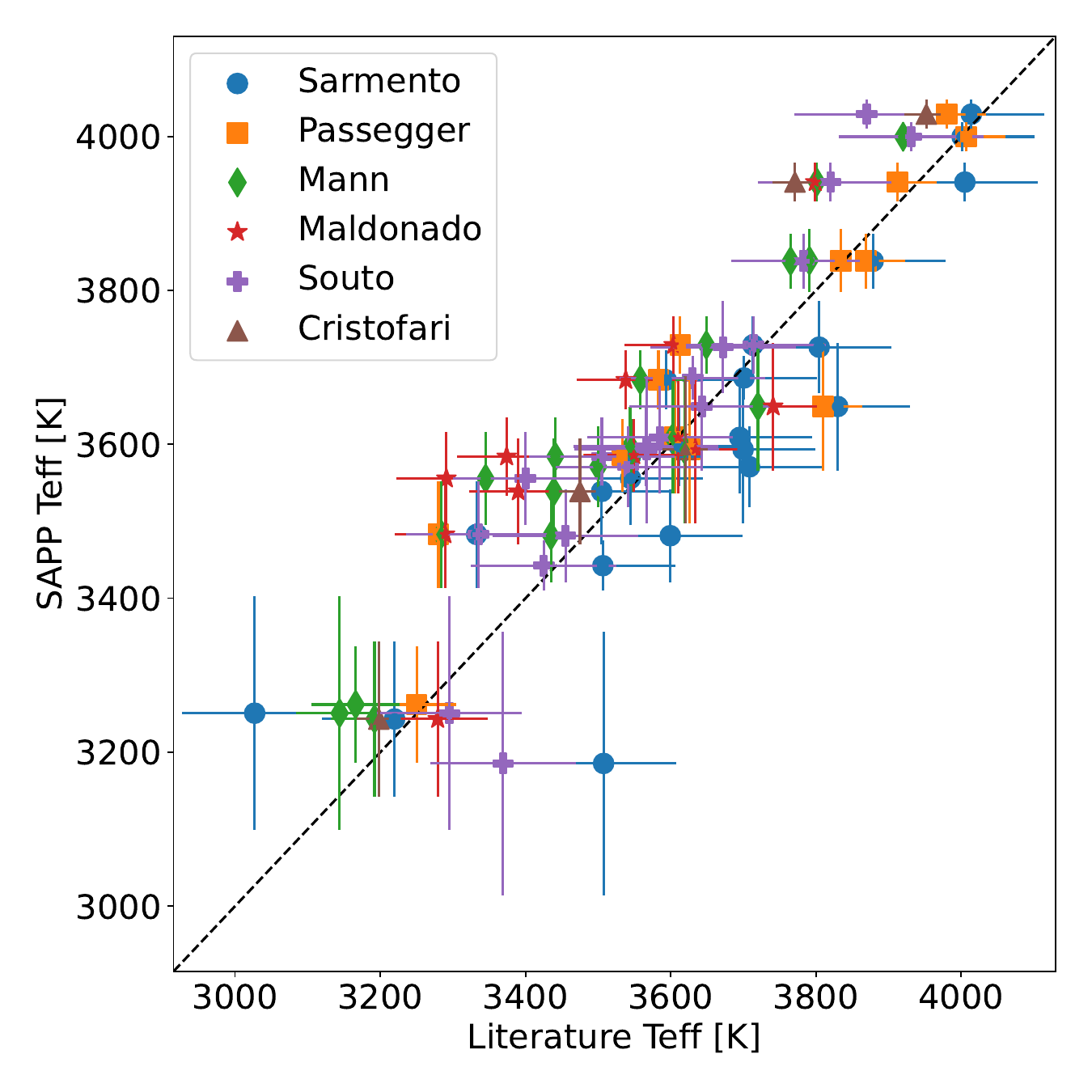}
   \includegraphics[width=6cm]{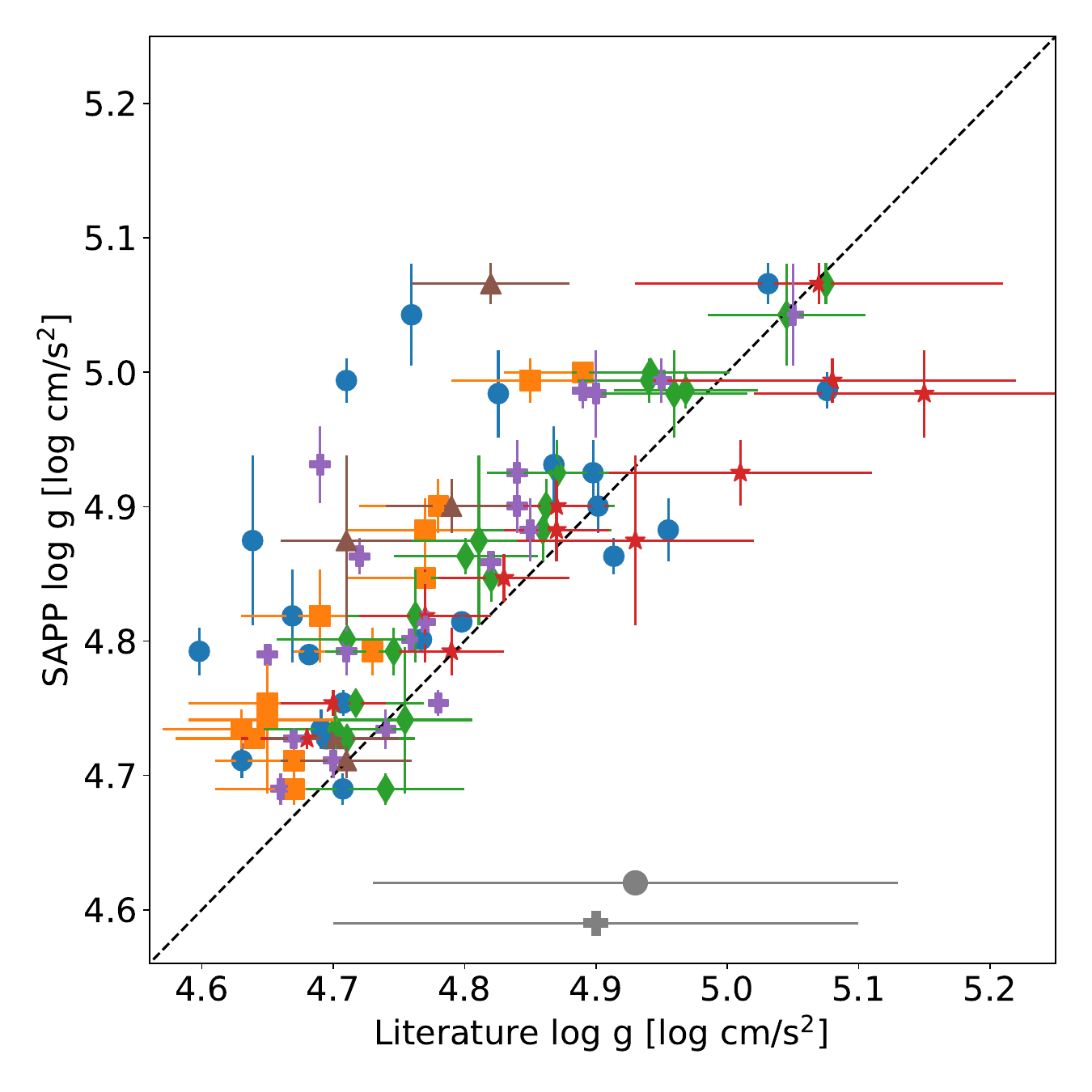}
   \includegraphics[width=6cm]{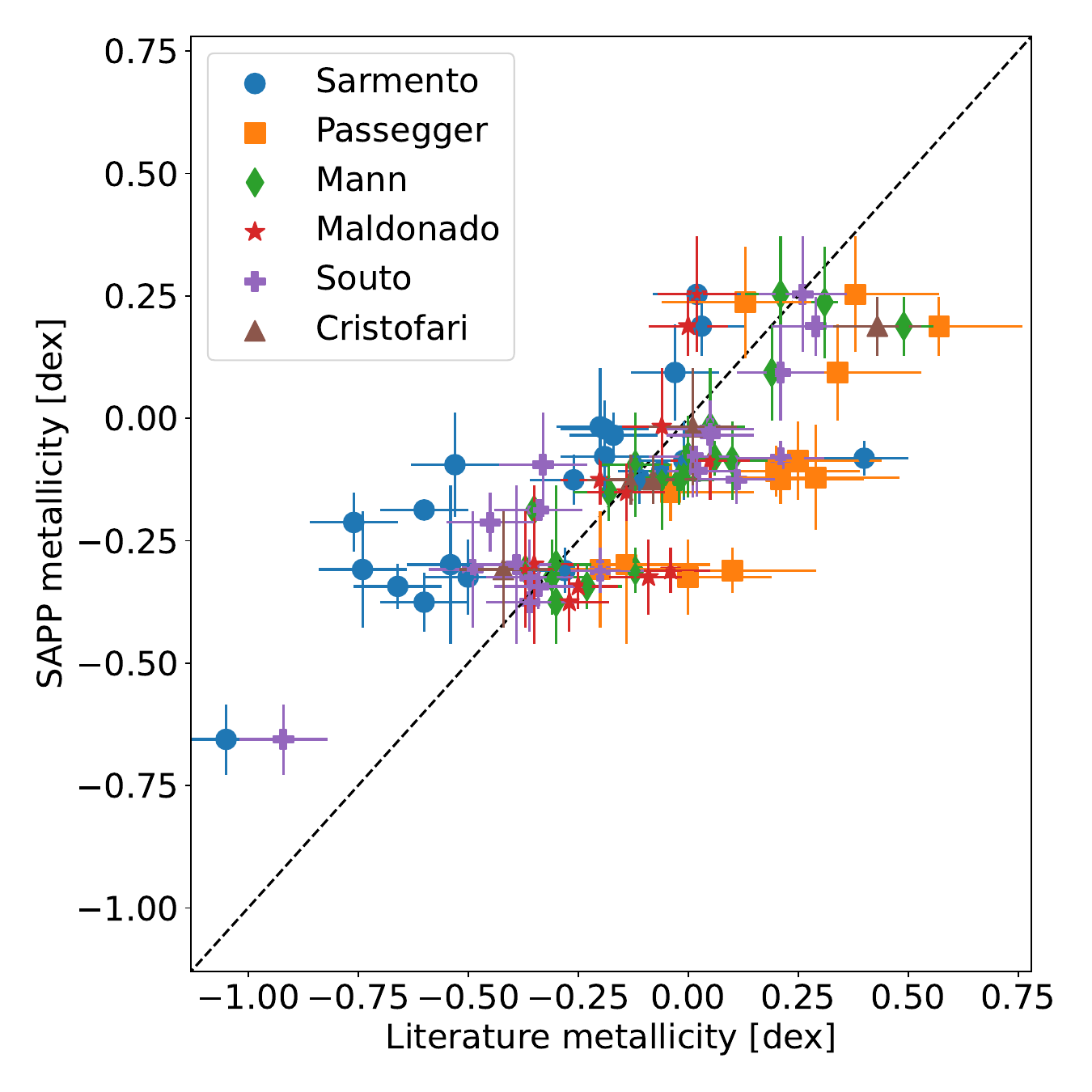}
   \caption{SAPP results compared with spectroscopic results from \citet{APOGEE_Sarmento2021,Pass2019,2015ApJMann,2020AAMaldonado,2022ApJSouto,2022MNRAS.516.3802C}.
   Values derived using the SAPP are shown on the vertical axis, and the literature values are shown on the horizontal axis.
   \textit{Left:} effective temperature.
   \textit{Middle:} surface gravity. The uncertainties for \citet{APOGEE_Sarmento2021} and \citet{2022ApJSouto} are represented at the bottom of the figure in grey. One star is located outside of the borders of the figure: LSPM~J1204+1728S for which the SAPP value is 4.86~dex and \citet{APOGEE_Sarmento2021} obtained 5.31~dex.
   \textit{Right:} metallicity. The black dashed line in all panels corresponds to the 1:1 ratio.
   }
   \label{Fig:Spec_param}
\end{figure*}

In the left panel of Fig.~\ref{Fig:Spec_param} we show the comparison for the effective temperature. For most stars the results agree within uncertainties. The $\teff$ values derived in the SAPP are slightly lower when comparing with \citet{APOGEE_Sarmento2021}, with a MAD of 83~K.
On the other hand, the SAPP derived on average higher $\teff$ values compared with \citet{2020AAMaldonado} and \citet{2015ApJMann}, with corresponding MADs of 120~K and 88~K, respectively.
We largely agree with \citet{2022ApJSouto}, \citet{Pass2019}, and \citet{2022MNRAS.516.3802C}, 
for which the corresponding MADs are 68~K, 62~K, and 76~K, respectively.
%
%
Thus, the MADs for the studies included in Fig.~\ref{Fig:Spec_param} are within 100~K in all cases, except for \citet{2020AAMaldonado}. As can be seen in Fig.~\ref{Fig:Spec_param} the uncertainties associated with the SAPP values are higher at lower effective temperatures. This could be due to the normalisation procedure having difficulties differentiating between noise and molecular lines for cooler stars. It could also just be a result of better $\teff$ diagnostics in the warmer M~dwarfs (stronger atomic lines and fewer molecular lines).

Two stars are apparent as outliers in $\teff$ in Fig.~\ref{Fig:Spec_param} (LSPM~J1204+1728S and GJ~777B).
For LSPM~J1204+1728S we derived a $\teff$ of 3185~K, while the values derived by \citet{2022ApJSouto} and \citet{APOGEE_Sarmento2021} are significantly higher (3369~K and 3507~K, respectively).
Our uncertainty for this star is high, and when visually inspecting the best-fit model we found that the agreement with the observations is poor. The lines in the observed spectra are significantly broader than in the best fit model which can be seen in Fig.~\ref{Fig:fast_rotator} in Appendix~\ref{append:fits}.
According to \citet{2018AJGilhool_Mdwarf_rot_APOGEE} this star is a fast rotator, with a $\vsini$ of about 17~kms$^{-1}$. The current version of the SAPP is not capable of fitting $\vsini$, and our result for this fast rotator is therefore not trustworthy. Future versions of the SAPP for M~dwarfs should also fit for rotational broadening of the spectral lines. This fast rotator only exists in the sample overlapping with \citet{APOGEE_Sarmento2021} and \citet{2022ApJSouto}. Recalculating the MAD without the star LSPM~J1204+1728S results in 72~K for \citet{APOGEE_Sarmento2021} and 62~K for \citet{2022ApJSouto}. 

For GJ~777B the $\teff$ derived by \citet[][3027~K]{APOGEE_Sarmento2021} is lower than ours (3251~K). On the other hand, the values derived by \citet[][3295~K]{2022ApJSouto} and \citet[][3144~K]{2015ApJMann} are in good agreement with ours. For this star the fit of our best model looks good. The SAPP uncertainty for this star is the highest in our sample (152~K), disregarding the fast rotator LSPM~J1204+1728S. In addition, GJ~777B has one of the lowest effective temperatures in our sample and the observed spectrum has a fairly low S/N (see Table~\ref{tab:Sample}). In our tests of the spectroscopic module without a line mask (i.e., using the complete spectral range) we found that for stars with low S/N and low $\teff$ we generally derived temperatures much higher than reference values. It is therefore a possibility that the line mask
is not fully appropriate for these types of spectra.
However, our derived $\teff$ is very similar to that from \citet{2022ApJSouto}. This star also requires further study.

\subsubsection{Surface gravity}
The middle panel of Fig.~\ref{Fig:Spec_param} shows a comparison for the surface gravity from the same studies as for the effective temperature. The uncertainties for \citet{APOGEE_Sarmento2021} and \citet{2022ApJSouto} are represented at the bottom of the figure as grey markers.
The largest deviations are found when comparing with \citet{APOGEE_Sarmento2021} who used a method relying largely on spectroscopy and did not constrain $\lgg$ with photometry, models and parallaxes as is done in the SAPP. The MAD is 0.11 dex. 
\citet{2022ApJSouto} used the oxygen abundance as a $\lgg$ indicator and their values also show a larger spread compared with the other studies in the figure. The MAD to our results is 0.07~dex. \citet{APOGEE_Sarmento2021} and \citet{2022ApJSouto} also quote the largest uncertainties, 0.2~dex in both cases.
Our derived $\lgg$ values agree fairly well with those of \citet{2015ApJMann}, \citet{2020AAMaldonado}, \citet{Pass2019}, and \citet{2022MNRAS.516.3802C}, 
with an apparent small systematic shift towards higher values in this work. The corresponding MADs are 0.04, 0.05, 0.09, and 0.11~dex, respectively. The differences between what was obtained from the SAPP and by \citet{Pass2019} or \citet{2022MNRAS.516.3802C} increase at higher surface gravities. The spread when comparing with \citet{2020AAMaldonado} also increases at higher surface gravities, as do the uncertainties from \citet{2020AAMaldonado}.

An outlier which is outside of the borders of the figure is the fast rotator mentioned above, LSPM~J1204+1728S. For this star, \citet{APOGEE_Sarmento2021} derived 5.31~dex using spectroscopy and we obtained 4.86~dex using photometry and evolutionary models. \citet{2022ApJSouto} obtained 4.82~dex for the same star. Excluding the fast rotator from the sample gives a MAD of 0.10~dex in comparison to \citet{APOGEE_Sarmento2021}. 

\subsubsection{Metallicity}
The right panel of Fig.~\ref{Fig:Spec_param} shows the metallicity derived using the SAPP compared with literature values.
Since M-dwarf metallicities are notoriously difficult to measure reliably (e.g. \citealt{APOGEE_Sarmento2021}, \citealt{2022AaA...658A.194P}, 
it is not surprising that this comparison shows significant deviations between independently-derived values.
The metallicity from the spectroscopic module of the modified SAPP is more confined to a region around solar metallicities compared with the other literature studies. Most of the metallicities from the SAPP lie between roughly $-0.5$~dex and +0.25~dex while the literature sample as a whole ranges from $-0.75$ to +0.4~dex (with a few additional values outside of these limits)\footnote{The apparent clumping of SAPP $\feh$ values in the horizontal direction is an effect of showing the results from several works for the same star. No clumping is seen when the comparison of SAPP results is done separately for each work.}. We note, however, that the values from \citet{APOGEE_Sarmento2021} cluster towards the lower limit and the \citet{Pass2019} values are found near the upper limit of the literature range.
The SAPP-derived metallicities are in general higher compared to \citet{APOGEE_Sarmento2021} and lower compared to \citet{Pass2019}. The corresponding MADs are 0.23~dex and 0.26~dex. The differences between our results and those of \citet{APOGEE_Sarmento2021} increase towards lower metallicites.
Our agreement is better with the studies by \citet{2015ApJMann}, \citet{2020AAMaldonado}, and \citet{2022ApJSouto}, and \citet{2022MNRAS.516.3802C}, 
for which the MADs are 0.10, 0.13, 0.13, and 0.09~dex, respectively.
The observed spread among the different studies is in line with the discussion in \citet{2022AaA...658A.194P}, 
who tested various different methods on the same CARMENES spectra. For some stars the results agreed well and for others differences of more than 0.5~dex were found (see their Figs.~2 and 3).

In our case, outliers in metallicity are mainly found in the comparison with \citet{APOGEE_Sarmento2021}.
For the star BD+00~549B we derived $-0.66$~dex, whereas \citet{APOGEE_Sarmento2021} derived $-1.05$~dex and \citet{2022ApJSouto} derived $-0.92$~dex. \citet{2018AJGilhool_Mdwarf_rot_APOGEE} obtained a metallcity of $-1.0$~dex, deriving the rotational velocity together with $\teff$, $\lgg$, and $\feh$ using a grid of template spectra compared to observed APOGEE spectra. We note that all studies we compare with for this particular star used some form of spectroscopic method to derive $\lgg$ while we used photometry and evolutionary models. It is a possibility that the difference is caused by difference in method and a degeneracy between $\lgg$ and $\feh$, but this should then apply to all stars in our sample. We note that this star is in a binary with an G star, see Sect.~\ref{subsec:comp_bin}.
Another outlier is the previously mentioned fast rotator LSPM~J1204+1728S, for which we obtained $-0.21$~dex with the SAPP, while \citet{APOGEE_Sarmento2021} and \citet{2022ApJSouto} obtained $-0.76$~dex and $-0.45$~dex, respectively. Excluding this fast rotator in the calculation of the MAD we obtain slightly lower values for both studies (0.22~dex and 0.123~dex, respectively).

Another outlier in comparison with \citet{APOGEE_Sarmento2021} is the star GJ~777B, for which we derived a metallicity of $-0.08$~dex with the SAPP and \citet{APOGEE_Sarmento2021} obtained a much higher value of +0.40~dex. \citet{2015ApJMann} derived +0.06~dex and \citet{2022ApJSouto} obtained +0.21~dex. This star was also mentioned as an outlier in effective temperature.
We note that GJ~777B is one of the coolest stars in the sample (spectral type M4.5, similar to GJ~324B and GJ~447), at the limits of validity of the usually employed atmospheric models and relations. This may contribute to the large spread in literature [Fe/H] values. This star is also part of a binary and discussed in Sect.~\ref{subsec:comp_bin}
We also note that for the stars in our sample overlapping with \citet{2020AAMaldonado} the authors did not derive a higher metallicity than +0.05~dex. However, they did obtain higher metallicities for other stars in their complete sample.


\subsection{Comparison with machine-learning techniques}

\begin{figure}
   \centering
   \includegraphics[width=8cm]{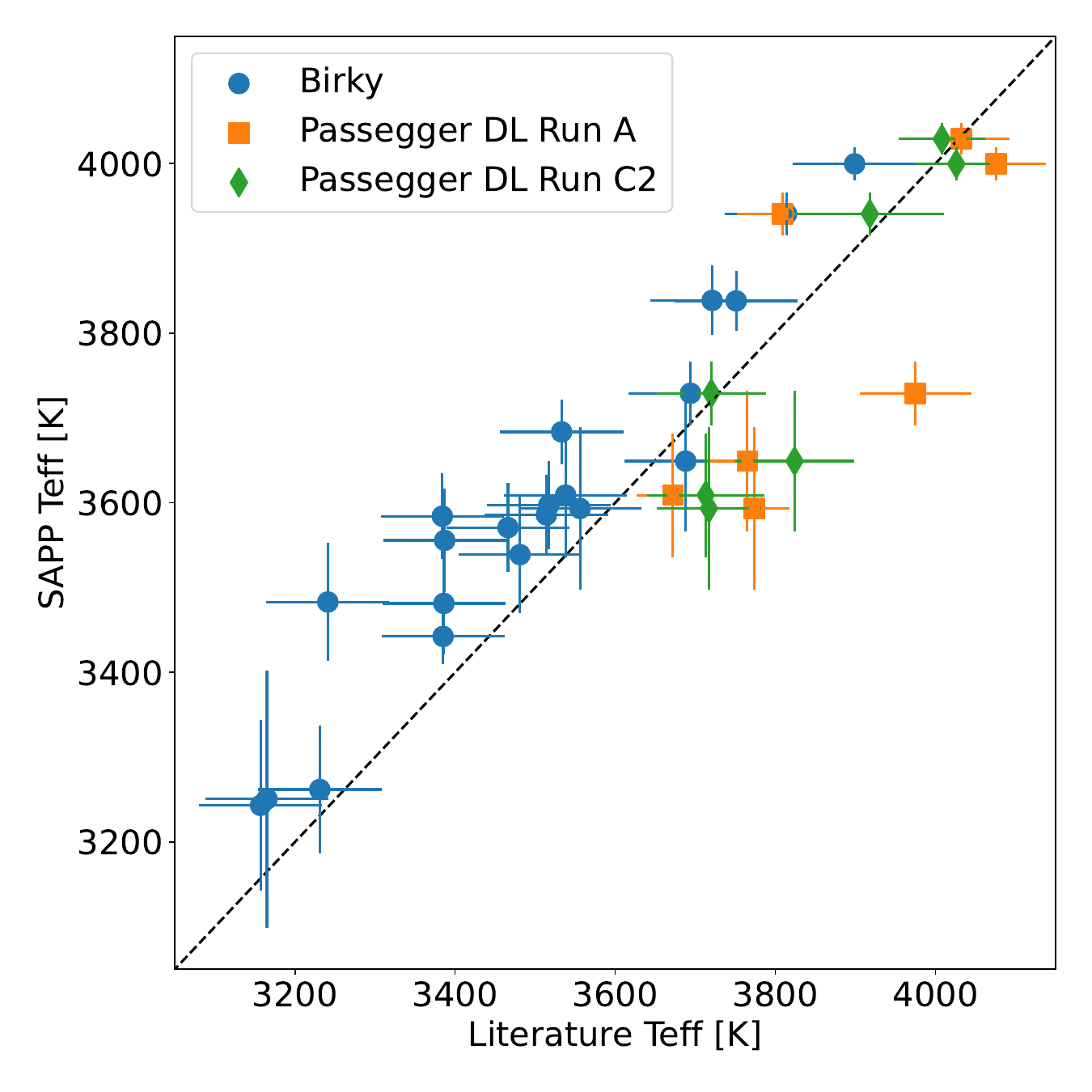}
   \includegraphics[width=8cm]{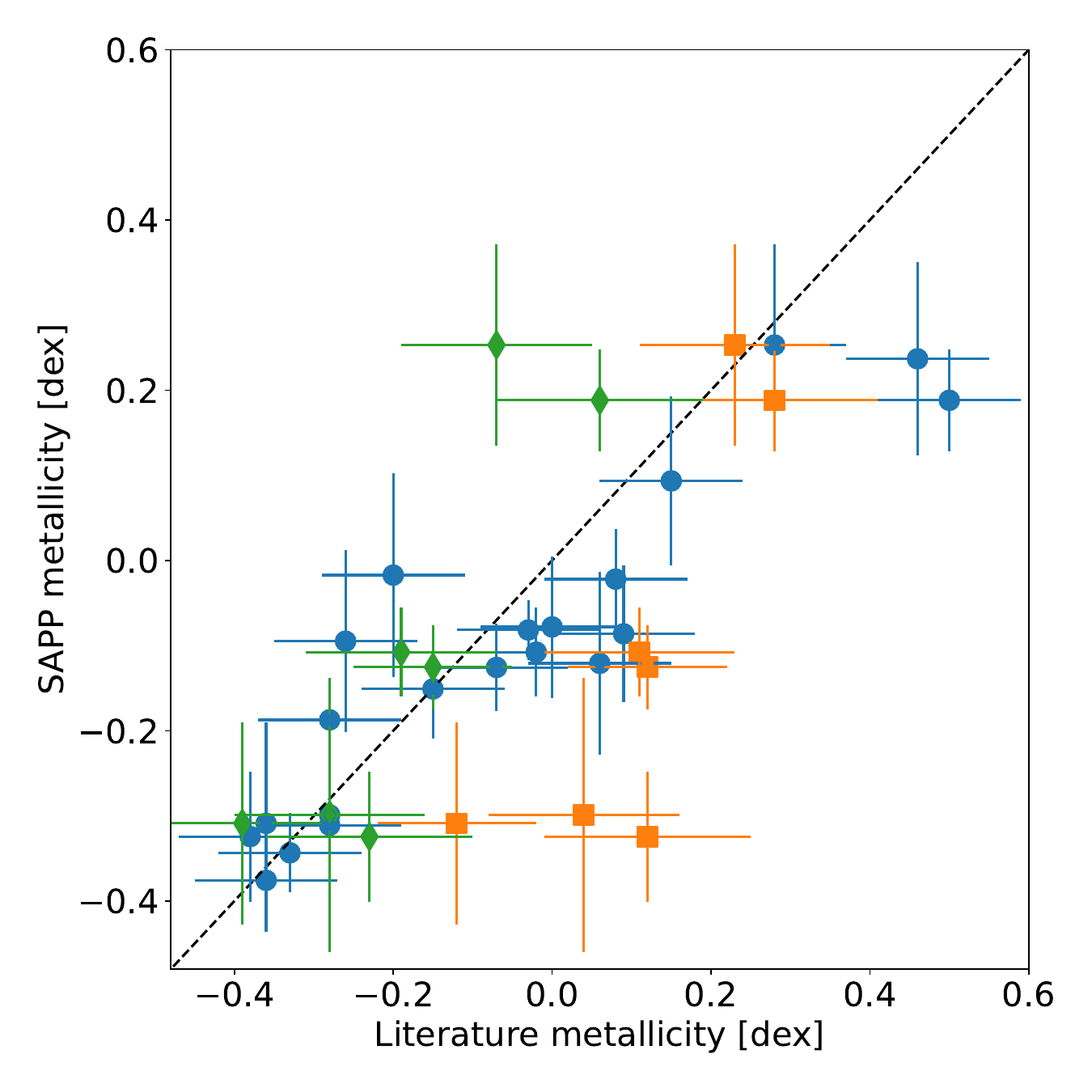}
   \caption{Comparing $\teff$ (top) and [Fe/H] (bottom) derived with the SAPP with the results based on machine-learning techniques from \citet{The_Cannon_Mdwarf_Birky2020ApJ} and \citet{2022AaA...658A.194P}. 
   The black dashed line corresponds to the 1:1 ratio.}
   \label{Fig:ML}
\end{figure}

A growing number of surveys are using machine learning methods to obtain stellar atmospheric parameters for large samples of stars. This includes the SAPP, in which we are using The Payne algorithm with an ANN trained on model spectra.
In this section we compare with the results of \citet{The_Cannon_Mdwarf_Birky2020ApJ}, who used another algorithm, The Cannon, trained on observed spectra, and with the results presented in \citet{2022AaA...658A.194P} 
based on a deep convolutional neural network trained on synthetic spectra (hereafter referred to as Deep Learning, DL).

The Cannon \citep{The_Cannon_Ness2015,The_Cannon_Casey2016} uses second-degree polynomial generative models trained on observed spectra and is therefore independent of stellar atmospheric models. Instead, it needs well-known benchmark stars. For M~dwarfs this can be a problem because of previously mentioned constraints on observing M~dwarfs.
\citet{The_Cannon_Mdwarf_Birky2020ApJ} used APOGEE spectra of a sample of well known M~dwarfs in their training of The Cannon and training labels from \citet{2011AJWest_Cannon_SptypeLabel} and \citet{2015ApJMann}. They subsequently applied their algorithm on other M~dwarfs from the APOGEE survey. The derived parameters were $\teff$ and $\feh$ as well as spectral type.
We use the ``Test'' values presented in their Table~2 for the comparison.

For the DL study, \citet{2020A&A...642A..22P} 
constructed a deep convolutional neural network architecture to produce neural network models trained on synthetic PHOENIX spectra. These were used to estimate $\teff$, $\lgg$, metallicity, and projected equatorial rotation velocity $\vsini$ for a sample of M~dwarfs from CARMENES spectra.
\citet{2022AaA...658A.194P} 
applied the same method to 18 well-studied M~dwarfs and compared the results to those obtained with several other methods, such as those mentioned in Sect.~\ref{subsec:result:comparison:spectroscopy}.
For the comparison we use the DL results from their Runs ``A'' and ``C2'', which differ in the spectral range used (a wavelength interval starting at 8800\,\AA\ for Run~A, and 35 wavelength windows distributed over the optical and J-band regions for Run~C2, see Table~2 in \citealt{2022AaA...658A.194P}).

A summary of these studies, including the instrument and the wavelength ranges used, is given in Table~\ref{tab:referenceStudies}.
Figure~\ref{Fig:ML} shows the literature results 
in comparison with ours 
for effective temperature and metallicity.
%
The effective temperature derived using the modified version of the SAPP is on average higher than that obtained by \citet{The_Cannon_Mdwarf_Birky2020ApJ}, with a MAD of 97~K.
There are no clear outliers, but we discuss here the three stars that are at the largest distance from the 1:1 ratio. Two of these are GJ~725A and B, which have been mentioned as outliers above. The $\teff$ values obtained with the SAPP for the (A, B) pair are (3584~K, 3556~K) and those obtained by \citet{The_Cannon_Mdwarf_Birky2020ApJ} are (3384~K, 3387~K). 
The third star is GJ~105B, for which we obtained 3483~K from the SAPP and \citet{The_Cannon_Mdwarf_Birky2020ApJ} obtained 3241~K.
Referring back to the works used for comparison in Sect.~\ref{subsec:result:comparison:spectroscopy}, we find that they also report consistently lower $\teff$ than that from the SAPP for this star, by about 200~K for \citet{Pass2019}, \citet{2015ApJMann}, and \citet{2020AAMaldonado}, and by about 150~K for \citet{APOGEE_Sarmento2021} and \citet{2022ApJSouto}. 
Regarding the $\teff$ values of the \citet{2022AaA...658A.194P} DL results, they agree well with the SAPP values, with a MAD of 117~K for Run~A and 69~K for Run~C2. At intermediate temperatures the SAPP values tend to be somewhat lower than the comparison values.

The SAPP derives on average somewhat higher metallicities in the lower metallicity range and somewhat lower metallicities in the higher metallicity range than \citet{The_Cannon_Mdwarf_Birky2020ApJ}, with a MAD of 0.09 dex for the whole range.
The SAPP metallicities agree well with those from \citet{2022AaA...658A.194P} DL Run~C2 at lower metallicity and with those from DL Run~A at higher metallicities, with a MAD of 0.22~dex for Run~A and 0.11~dex for Run~C2 over the whole range.
The outliers at the high-metallicity end are GJ~205 (0.19, 0.50, 0.06~dex), GJ~324B (0.24, 0.46~dex, none), and GJ~880 (0.25, 0.28, $-0.07$~dex), with values derived by the SAPP, \citet{The_Cannon_Mdwarf_Birky2020ApJ}, and \citet{2022AaA...658A.194P} DL Run~C2, respectively, given in parentheses.
For GJ~205, the works used for comparison in Fig.~\ref{Fig:Spec_param} report metallicities between 0.00 and 0.57~dex, implicating a need for further investigation of this star. 
For GJ~324B, \citet{Pass2019} obtained 0.13~dex and \citet{2015ApJMann} obtained 0.31~dex. This star is in a binary with an early K~star (GJ~324A), for which \citet{2018Montes_FGKmetal} derived 0.29~dex. It appears that our derived metallicity is closer to the metallicity of the primary star in the binary than the metallicity of \citet{The_Cannon_Mdwarf_Birky2020ApJ} or \citet{2022AaA...658A.194P} DL Run~C2.
We note that \citet{2022AaA...658A.194P} concluded that the results from their DL Run~A were most consistent with those of other methods used in the same work and in the literature. Here, we find a slight preference for DL Run~C2 for the limited sample in common, for both $\teff$ and metallicity.



\subsection{Comparison of binary components}
\label{subsec:comp_bin}

   \begin{table*}[ht]
   \caption{Parameters of FGK-type primary stars with an M~dwarf secondary.}
   \label{tab:Binary_param}
       \centering
       \begin{tabular}{l c c c l c l}
       \noalign{\smallskip}
        \hline
        \noalign{\smallskip}
        Primary  &  $\teff$/K  &  $\lggu$  &  [Fe/H]  &  Sp. type & Ref. & Secondary\\
        \noalign{\smallskip}
        \hline
        \noalign{\smallskip}
HD24421         & 6108$\pm$26 & 4.30$\pm$0.05 & -0.32$\pm$0.02 & F5V    & 1 & LSPMJ0355+5214 \\
GJ297.2A        & 6414$\pm$51 & 4.65$\pm$0.09 & -0.01$\pm$0.03 & F6.5V  & 2 & GJ297.2B \\
GJ324A          & 5299$\pm$58 & 4.35$\pm$0.13 &  0.29$\pm$0.04 & K0IV-V & 3 & GJ324B \\
GJ3194          & 5644$\pm$21 & 4.36$\pm$0.05 & -0.30$\pm$0.02 & G1.5V  & 3 & GJ3195 \\
BD+00549        & 5319$\pm$46 & 4.48$\pm$0.12 & -0.88$\pm$0.04 & G5     & 4 & BD+00549B \\
BD-064756       & 4494$\pm$44 & 4.69$\pm$0.06 & -0.01$\pm$0.03 & K5-V   & 2 & BD-064756B \\
HD122972        & 5523$\pm$22 & 4.36$\pm$0.07 & -0.01$\pm$0.02 & G6V    & 5 & Ross799 \\
GJ777A          & 5578$\pm$22 & 4.27$\pm$0.05 &  0.21$\pm$0.02 & G7IV-V & 2 & GJ777B \\
GJ211           & 5292$\pm$32 & 4.38$\pm$0.09 &  0.04$\pm$0.02 & K0V    & 3 & GJ212 \\
LSPMJ1204+1728N & 5282$\pm$34 & 4.41$\pm$0.09 & -0.18$\pm$0.02 & G9V    & 6 & LSPMJ1204+1728S \\
        \noalign{\smallskip}
        \hline
       \end{tabular}
   \tablefoot{
   All parameters for the primary components in the table were obtained from \citet{2018Montes_FGKmetal} except for BD$-$06~4756 for which \citet{2013AJ....145...52M} was used.
   The parameters for the K-type primaries with APOGEE spectra that were analysed by the SAPP (GJ~105A, GJ~338A) are given in Table~\ref{tab:Result_analysis}. 
   Column ``Ref.'' gives the reference for the spectral type.
   \citet{2006AJ....132..161G} used the system of \citet{1989ApJS...71..245K} for the spectral types.
   }
   \tablebib{(1) \citet{1993yCat.3135....0C}; (2) \citet{2006AJ....132..161G}; (3) \citet{2003AJ....126.2048G}; (4) \citet{1984AJ.....89..702L}; (5) \citet{1999MSS...C05....0H}; (6) \citet{2018MNRAS.481.3244G}}
   \end{table*}

\begin{figure}
   \centering
   \includegraphics[width=9cm]{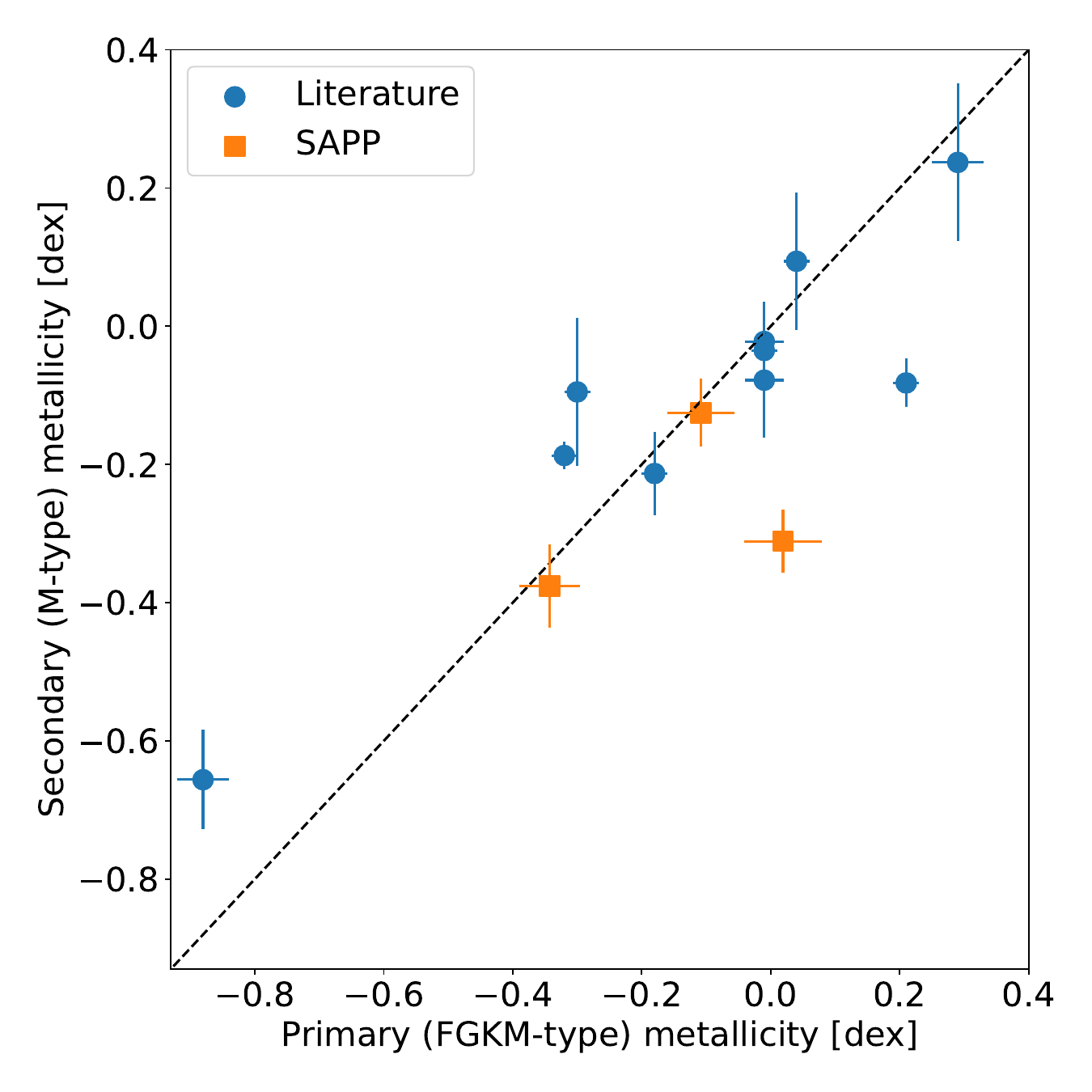}
   \caption{Comparing the derived $\feh$ from the SAPP for M~dwarf secondary components in a binary (y-axis, Table~\ref{tab:Result_analysis}) with the metallicity of the primary (x-axis, Table~\ref{tab:Binary_param}). The blue circles show binaries with an FGK-type primary and an M~dwarf secondary. The orange squares show binaries with a KM-type primary and an M~dwarf secondary. All stars in these three systems were analysed with the SAPP using APOGEE spectra. The black dashed line corresponds to the 1:1 ratio.}
   \label{Fig:Binaries_met}
\end{figure}

Our sample includes M~dwarfs in 13 binary systems with other M~dwarfs as well as with FGK-type stars.
Analysing both stars in a binary allows one to verify the internal consistency, as any metallicity difference between the two components gives an indication of systematics inherent in the analysis method. For the cases where the two components are very different in terms of spectral type, this holds assuming that differential diffusion effects can be neglected.
The stellar parameters of the FGK-type primaries are given in Table~\ref{tab:Binary_param} and were taken from the literature, except for two late K dwarfs which were included in the analysis in this work.
%
%
We use metallicities from \citet{2018Montes_FGKmetal}, who analysed optical spectra, and from \citet{2013AJ....145...52M} based on moderate resolution visible and infrared spectra. 
When the primary star was within the parameter range of the modified version of the SAPP and an APOGEE spectrum was available it was analysed with the SAPP. This was the case for GJ~338A and B and GJ~725A and B, in which both stars are either late K~dwarfs or early M~dwarfs, and for GJ~105A and B, which are an earlier K~dwarf and an M~dwarf.

The results are compared in Fig.~\ref{Fig:Binaries_met} where the metallicity of the primary is shown on the horizontal axis and the metallicity of the secondary on the vertical axis. Different symbols distinguish the binaries for which the metallicity comparison data were taken from the literature from the binaries where both stars were analysed with the SAPP. The stars largely follow the 1:1 ratio with a small spread. The majority of the binaries have a difference in metallicity between the primary and secondary smaller than 0.15~dex.

The two systems which deviate most from the 1:1 line in the comparison with literature values are the (A, B) pairs (BD+00~549, BD+00~549B) where the primary has a metallicity of $-0.88$~dex and the SAPP derived $-0.66$~dex for the secondary, and (GJ~777A, GJ~777B) with metallicities of (0.21~dex, $-0.08$~dex), respectively. Both of the secondaries have previously been mentioned as outliers with regard to our derived metallicity.
An additional outlier when comparing with FGK-type primaries is the pair (GJ~3194, GJ~3195), where the literature value for the primary is $-0.30$~dex, and we derived $-0.10$~dex for the secondary. For comparison, the values derived by \citet{2022ApJSouto} and \citet{APOGEE_Sarmento2021} for the secondary are $-0.33$~dex and $-0.53$~dex. We note that our effective temperature of 3570~K deviates from that by \citet{APOGEE_Sarmento2021} of 3708~K for this star. \citet{2015ApJMann} derived $-0.12$~dex for the metallicity, which is much closer to our value. 
The outlier in the sample where both components were analysed with the SAPP is the binary GJ~105 A/B. For the primary, the SAPP derived 0.02~dex, and for the secondary $-0.31$~dex was obtained. The pipeline is optimised for analysing M~dwarfs and the result for the K-type star GJ~105A can therefore be considered to be unreliable. \citet{2018Montes_FGKmetal} quote a metallicity of $-0.20$~dex for the primary component, which is closer to the value obtained for the secondary with the SAPP.
We can conclude that our method is applicable at least up to K7 stars in terms of effective temperature (the case of GJ~338A).


\subsection{Estimated overall uncertainties}
\label{subsec:result:uncertainties}
Summarising the results presented in the previous sections, we find a systematic offset in $\teff$ compared to interferometric values of about 100~K. Based on this offset and the mean absolute differences calculated with respect to other studies, we estimate our overall uncertainty in $\teff$ to be 100~K.
When comparing with literature values of $\lgg$ the SAPP surface gravities seem to be about 0.1~dex higher. At lower surface gravities this offset is lower. We therefore estimate the general uncertainty in surface gravity derived by the SAPP to be 0.1~dex.
Regarding the metallicity, we find differences between binary components of up to 0.2~dex, but for most stars the difference is below 0.15~dex. The MADs calculated with respect to other studies are between 0.1 and 0.26~dex. The higher absolute differences occur when comparing with \citet{APOGEE_Sarmento2021} and \citet{Pass2019}. The median of the MADs is 0.13~dex and the mean is 0.16~dex. We therefore estimate the overall uncertainty of the SAPP-derived metallicity to be 0.15~dex.
This estimate can be viewed in the light of the investigation of uncertainties related to abundance determinations in APOGEE M~dwarf spectra presented in \citet[Appendix~A]{2024ApJ...973...90M}. 
Based on simulated spectra, uncertainties were calculated as a function of S/N, $\teff$, and shifts in the pseudo-continuum level. For most elements\footnote{Exceptions were Si and Na. We note that our line mask contains several Si features, but no Na lines.}, typical uncertainties were around 0.05~dex for S/N $\gtrsim$ 100, reaching $\sim$0.15~dex for lower S/N at the cool end. Typical abundance uncertainties due to continuum shifts of 1\% were 0.1~dex.
Future development of the SAPP for M~dwarfs will aim to improve the precision and accuracy for the derivation of the stellar parameters.


\section{Future developments}
\label{sec:future}

As shown in the previous sections the results of the modified SAPP for M~dwarfs look promising. However, we have identified possible areas of future development that should lead to an improved accuracy and precision of the derived stellar parameters. The capabilities of the pipeline should also be extended to enable the derivation of further parameters, such as rotational velocity and abundances of individual chemical elements.

\subsection{Line lists and line mask}
It is important to update the line lists used to calculate the grid of synthetic spectra used to train The Payne ANN with the latest atomic and molecular data. In addition, the APOGEE DR16 line list was created for the entire APOGEE survey. It is desirable to design a line list optimised for M~dwarfs, which should improve the results of the SAPP.
Another factor concerning the synthetic spectra is hyperfine structure splitting, which can severely affect lines in M~dwarf spectra, as shown for example by \citet{2021YutongShan} for vanadium lines. However, hyperfine structure of V and other elements is included in the APOGEE DR16 line list \citep{APOGEE_DR16_linelist_Smith2021}, and should therefore not be of much concern. 

The fitting procedure in the spectroscopic module is applied to selected spectral ranges for which the models are deemed to be most reliable (see Fig.~\ref{Fig:GJ880_spec}). The line mask used in the SAPP version presented here is based on published work done in another context \citep{APOGEE_Sarmento2021} and needs to be optimised for the \plato\ pipeline. Furthermore, in its current preliminary setup the spectroscopic module can only determine $\teff$ and metallicity for M~dwarfs. Spectroscopic diagnostics for $\lgg$ and abundances of individual elements need to be identified. In this regard future work could investigate whether using different line masks when fitting $\lgg$, $\teff$ and metallicity could improve the results.

It may also be worthwhile to explore the extension of the analysis to other NIR regions, for example the J-band at 1.1 to 1.4~$\mu$m. An example for an abundance analysis of several early- and mid-M~dwarfs in the J-band using CARMENES spectra is given by \citet{2020PASJ...72..102I}. 
This region combines the advantages of containing maximum flux and minimal molecular absorption by water for M~dwarfs. In practice, this would require an extension of the line list, the synthetic spectra grid, and the training of the ANN, as well as an adaptation of the normalisation procedure. 

\subsection{Non-LTE}
\label{sec:future:nlte}
In collision-dominated atmospheres LTE is assumed.
This may be expected to be the case for the atmospheres of M~dwarfs because of their high density. However, recent studies have shown departures from LTE in M~dwarfs for a number of elements.
\citet{Hauschildt_1997ApJ...488..428H} investigated non-LTE effects on titanium lines for M type stars, both giants and dwarfs. They found that titanium lines in M~dwarfs are stronger in non-LTE compared to LTE and that this effect decreases with effective temperature. 
\citet{2020AbiaSrRb} studied the elements Rb and Sr. They found an average non-LTE abundance correction of $-0.15$~dex for Rb lines for a range of stellar parameters covering late K~dwarfs and early M~dwarfs. They also derived abundance corrections for Sr lines varying between $-0.28$ and $-0.13$~dex. The abundance correction decreases with increasing temperature for both elements in this study.
\citet{Olander2021} showed that non-LTE effects can cause abundance differences of up to 0.2~dex for potassium lines.

Therefore, it seems clear that non-LTE effects for atomic lines need to be taken into account when analysing M~dwarfs. 
Future training grids to be used in the SAPP for M~dwarfs should be generated using non-LTE departure coefficents, as already done for FGK-type stars in \citet{Gent_2022A&A}.
We note that the diagnostic lines in the spectra used in this work include a number of molecular features (mainly from OH, CN, NO, SiH). However, non-LTE studies of molecular line formation are rare, and they have so far focused on CO, CH, and water in the Sun and cool giants (see e.g. Sect.~5 in the review by \citealt{2016A&ARv..24....9B} or Sect.~2.5.2 in \citealt{2024arXiv240100697L}).
In summary, there is a need for more research regarding non-LTE in M~dwarfs.

\subsection{Rotation}
\label{sec:future:rotation}
One star in our sample for which the SAPP tends to give problematic results is LSPM~J1204+1728S, which is a fast rotator. The discrepancy between the parameters derived by the SAPP and given in the literature, and the poor fit between the model and observed spectra for this star show that the current version of the SAPP for M~dwarfs is not applicable to stars with high rotation velocities. Future developments of the pipeline in regards to M~dwarfs should include the rotational velocity as a fitting parameter. In addition, faster rotating stars also tend to have stronger magnetic fields \citep[e.g.][]{2012AJ....143...93R}. 
If fast rotators are to be analysed care must be taken to avoid magnetically sensitive lines.

\subsection{Magnetic fields}
The presence of a magnetic field broadens spectral lines through the Zeeman effect. In addition, equivalent widths of strong spectral lines with many Zeeman components are increased due to the effect of magnetic intensification. Both of these effects were discussed in the context of M~dwarf studies by \citet{2021Kochukhov} and employed to determine mean magnetic fields of hundreds of M~dwarfs in a series of studies based on CARMENES spectra \citep{Shulyak2019,Reiners2022}. These investigations revealed fields in the range from a few hundred G in slowly rotating inactive M~dwarfs all the way to 6--8~kG in the most active stars. Strong fields in the kG-range would certainly have an impact on the fitting procedure in the spectroscopic module. We note that most stars in the sample used in this work are inactive stars with 200--600~G fields \citep{Reiners2022}, which do not produce noticeable line distortions or intensity changes at the resolution of the APOGEE spectra. However, for future analysis with the M~dwarf version of the SAPP where more stars are targeted magnetic fields need to be taken into consideration, for example by avoiding magnetically sensitive lines in the line mask. 


\section{Conclusions}
\label{sec:conclusion}
In preparation for the launch of the \plato\ telescope a prototype pipeline that derives stellar parameters for FGK~stars has been developed, the SAPP. It uses Bayesian inference to combine results from spectroscopy, photometry, and asteroseismology in order to obtain reliable stellar parameters such as effective temperature, surface gravity, metallicity, and abundances. In this article we present a modified version of the pipeline that is capable of analysing M~dwarf spectra in the H-band. We focus on the spectroscopic and photometric parts of the code and leave the full Bayesian analysis to future work. We used the pipeline to derive the three main parameters $\teff$, $\lgg$, and $\feh$ and assessed its performance on a sample of reference stars with APOGEE spectra and independent parameter determinations from the literature. Other parameters are left to future work.

The surface gravity is constrained using photometry and stellar interior models. We implemented a new grid of stellar evolutionary models specifically calculated for the analysis of M~dwarfs in the framework of the BaSTI library. Observed magnitudes in different photometric bands are used together with distances to calculate absolute magnitudes. 
These are compared to synthetic absolute magnitudes calculated from the evolutionary models. Probability distribution functions are computed to estimate the most probable stellar parameters, and the $\lgg$ value is passed on to the spectroscopic module.

The spectroscopic module of the SAPP fits synthetic spectra based on a model grid to observed spectra, in order to derive atmospheric parameters. The code does not use the synthetic spectra from the grid directly, but uses a fast model-reconstruction technique, based on the machine learning algorithm ``The Payne''. We used Turbospectrum, MARCS atmospheric models, and the APOGEE DR16 line list together with the water line list by \citet{2018MNRASPolyansky} to generate a grid of synthetic spectra in the H-band covering the M~dwarf range of atmospheric parameters, which was used for training an ANN used by The Payne.
In preparation for the fitting process the observed spectra need to be normalised. The built-in normalisation procedure in the SAPP was adjusted in order to take the pseudo-continuum encountered in M~dwarf spectra into account. We generated a grid of synthetic spectra, using the same setup as for generating the training grid. A second degree polynomial was then fitted to the upper envelope of the flux as a function of wavelength for each synthetic spectrum, resulting in a set of polynomials for different effective temperatures. The code starts with a fit to the observed spectrum that has been normalised by the original SAPP normalisation routine. The resulting initial stellar parameters are used to find the corresponding polynomial and the continuum is adjusted accordingly. The adjusted spectrum is then used in a new fit. This procedure is repeated iteratively to convergence.
The fitting procedure is applied to selected spectral ranges within a line mask considered to be suitable for M~dwarfs.

In summary, the adaptations of the SAPP comprise the following calculations and procedures specific to M~dwarfs: the evolutionary models and synthetic photometry, the synthetic spectra grid in the NIR, the trained ANN model, the normalisation of the observed spectra, and the line mask. 
The results derived with the modified SAPP for our sample of reference M~dwarfs agree in general well with the results from a number of literature studies that applied a variety of methods.
Our derived effective temperatures seem to be about 100~K higher than those calculated from interferometric angular diameters and bolometric fluxes.
For the surface gravity, the SAPP produces values consistent with those derived by other studies based on photometry, for example \citet{2015ApJMann}.
To assess the metallicity performance we compared the metallicities of the component stars in binary systems with an FGK~type primary and an M~dwarf secondary. We find an agreement within about 0.15~dex.
In future work, we plan to analyse the primary stars of the binary sample with the FGK-version of the SAPP. This will allow us to compare the performance of the two channels of the SAPP, aiming for a smooth transition in the overlap region of spectral types (K dwarfs).

The performance of the pipeline for M~dwarfs is expected to improve following future development.
New grids of model spectra will be calculated, updated line lists implemented, and the effects of rotation and non-LTE taken into account. The setup of the spectroscopic module will be refined to enable diagnostics for surface gravity and abundances of individual elements, and to mitigate the effects of magnetic fields.
The photometric module will be extended to longer-wavelength bands and to higher metallicities.
Finally, the full Bayesian inference scheme will be implemented, combining probability distribution functions from both the spectroscopic and the photometric module to deliver the most reliable properties for the \plato\ M~dwarfs sample.


\begin{acknowledgements}
We thank Andrew Mann, Vera Passegger, and Denis Mourard for useful discussions.
T.O., U.H., O.K., and N.J.M. acknowledge support from the Swedish National Space Agency (SNSA/Rymdstyrelsen). O.K. also acknowledges support by the Swedish Research Council (grant agreement no. 2019-03548) and the Royal Swedish Academy of Sciences.
M.B. is supported through the Lise Meitner grant from the Max Planck Society. This project has received funding from the European Research Council (ERC) under the European Unions Horizon 2020 research and innovation programme (Grant agreement No. 949173).
E.M. acknowledges support by the Collaborative Research centre SFB 881, Heidelberg University, of the Deutsche Forschungsgemeinschaft (DFG, German Research Foundation).
S.C. has been funded by the European Union – NextGenerationEU" RRF M4C2 1.1  n: 2022HY2NSX. "CHRONOS: adjusting the clock(s) to unveil the CHRONO-chemo-dynamical Structure of the Galaxy” (PI: S. Cassisi), and by INAF Theory grant "Lasting" (PI: S. Cassisi).
T.M. acknowledges financial support from Belspo for contract PRODEX PLATO mission development.
H.S.W. acknowledges support from the Carlsberg Foundation's Semper Ardens grant (``FIRSTATMO''; PI: A. Johansen).
D.S. thanks the National Council for Scientific and Technological Development -- CNPq.
B.R.-A. acknowledges funding support from the ANID Basal project FB210003.
E.D.M. acknowledges the support by the Ram\'on y Cajal contract
RyC2022-035854-I funded by the Spanish MICIU/AEI/10.13039/501100011033
and by ESF+.
This research was partially supported by the project AI4Research at Uppsala University.
Funding for the Sloan Digital Sky Survey IV has been provided by the Alfred P. Sloan Foundation, the U.S. Department of Energy Office of Science, and the Participating Institutions.
SDSS-IV acknowledges support and resources from the Center for High Performance Computing at the University of Utah. The SDSS website is www.sdss4.org.
SDSS-IV is managed by the Astrophysical Research Consortium for the Participating Institutions of the SDSS Collaboration including the Brazilian Participation Group, the Carnegie Institution for Science, Carnegie Mellon University, Center for Astrophysics / Harvard \& Smithsonian, the Chilean Participation Group, the French Participation Group, Instituto de Astrof\'isica de Canarias, The Johns Hopkins University, Kavli Institute for the Physics and Mathematics of the Universe (IPMU) / University of Tokyo, the Korean Participation Group, Lawrence Berkeley National Laboratory, Leibniz Institut f\"ur Astrophysik Potsdam (AIP), Max-Planck-Institut f\"ur Astronomie (MPIA Heidelberg), Max-Planck-Institut f\"ur Astrophysik (MPA Garching), Max-Planck-Institut f\"ur Extraterrestrische Physik (MPE), National Astronomical Observatories of China, New Mexico State University, New York University, University of Notre Dame, Observat\'ario Nacional / MCTI, The Ohio State University, Pennsylvania State University, Shanghai Astronomical Observatory, United Kingdom Participation Group, Universidad Nacional Aut\'onoma de M\'exico, University of Arizona, University of Colorado Boulder, University of Oxford, University of Portsmouth, University of Utah, University of Virginia, University of Washington, University of Wisconsin, Vanderbilt University, and Yale University.
This work has made use of data from the European Space Agency (ESA) mission
{\it Gaia} (\url{https://www.cosmos.esa.int/gaia}), processed by the {\it Gaia}
Data Processing and Analysis Consortium (DPAC,
\url{https://www.cosmos.esa.int/web/gaia/dpac/consortium}). Funding for the DPAC
has been provided by national institutions, in particular the institutions
participating in the {\it Gaia} Multilateral Agreement.
\end{acknowledgements}

\bibliographystyle{aa}


\begin{appendix}

\section{Validation of the generative ANN model}
\label{append:validation}

\begin{figure}
   \centering
   \includegraphics[width=8cm]{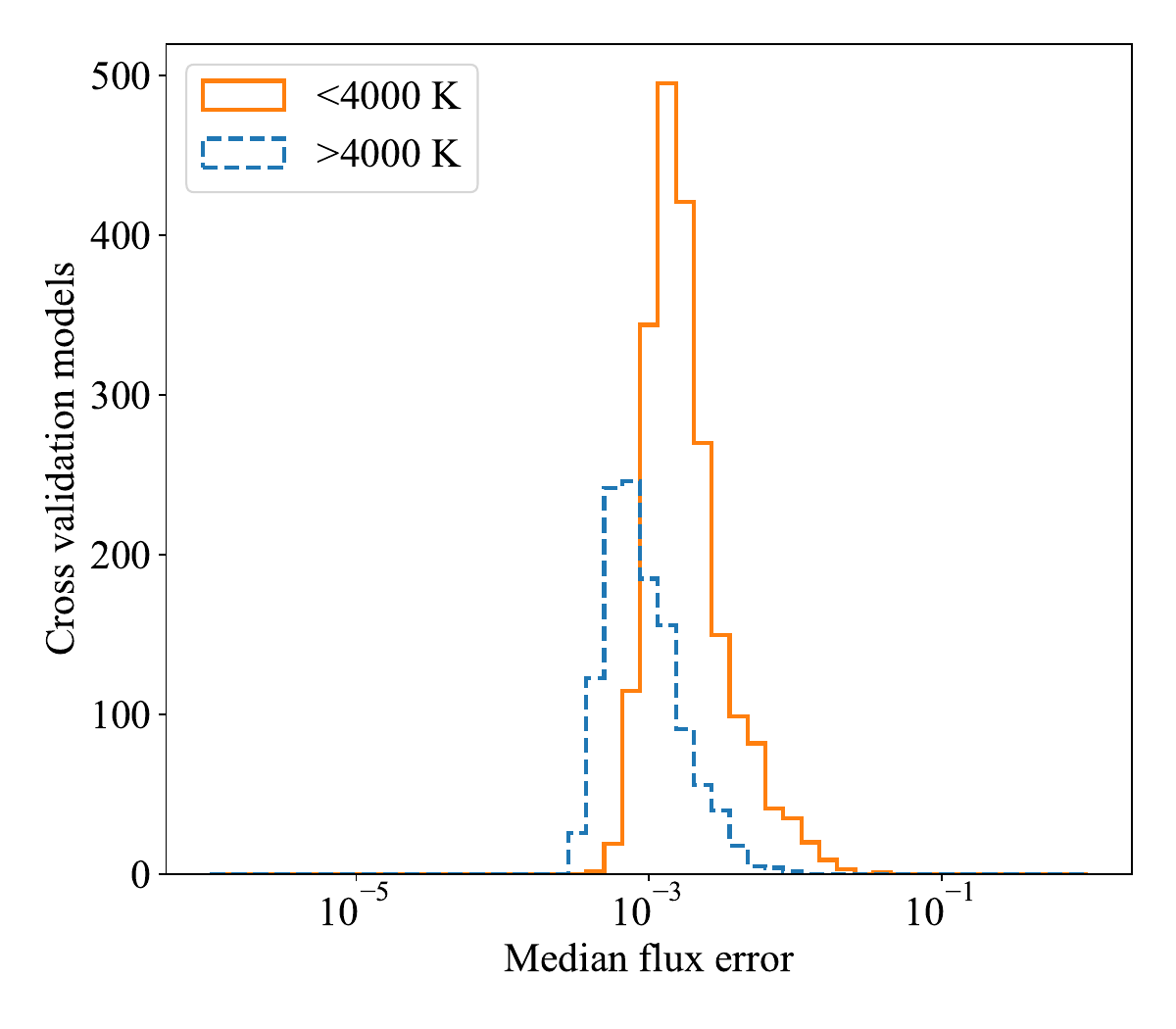}
   \includegraphics[width=8cm]{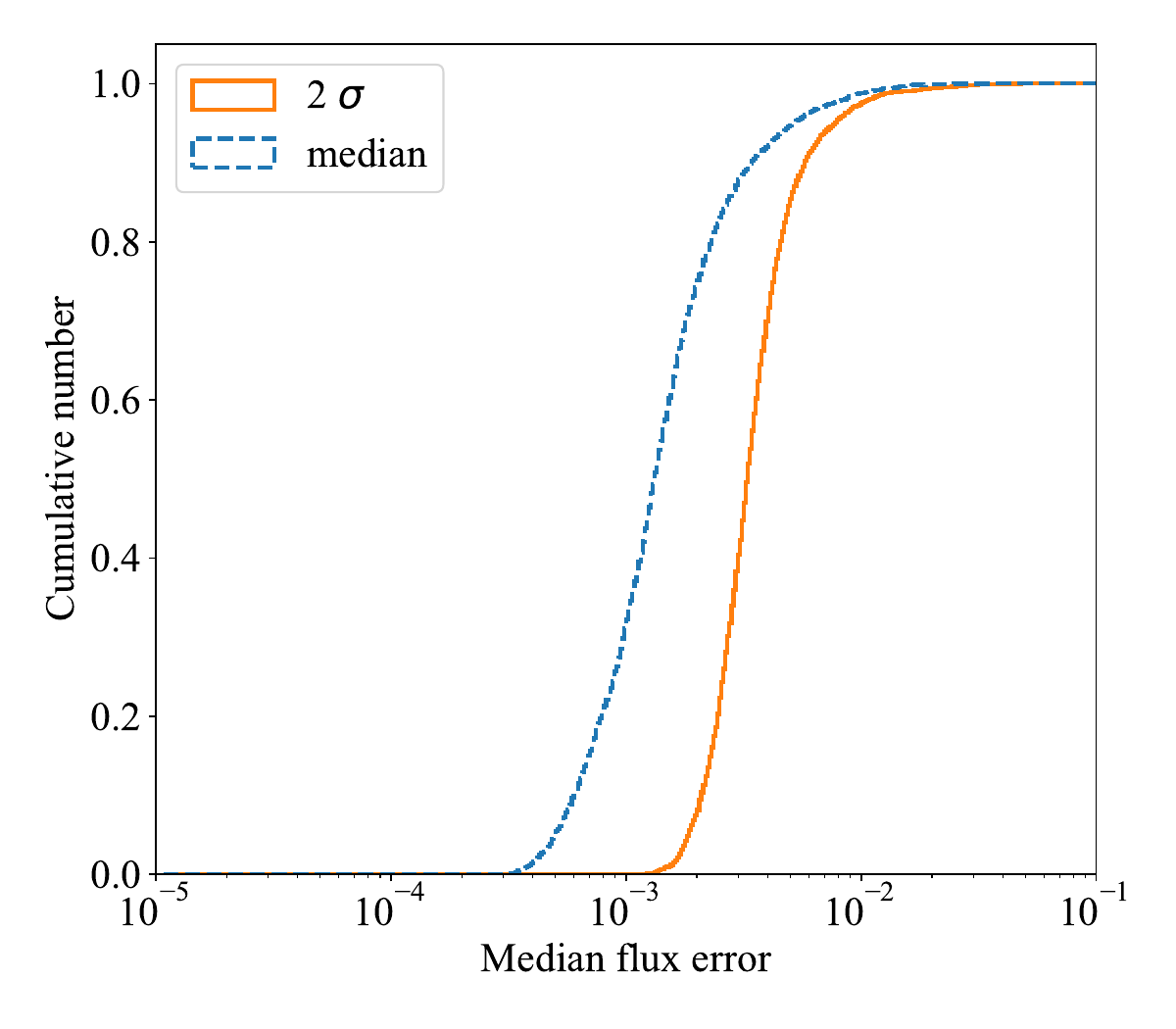}
   \caption{Top: distribution of median flux error for 3300 models from the validation set of The Payne's ANN, bottom: cumulative distribution.}
   \label{Fig:histogram}
\end{figure}

As mentioned in Sect.~\ref{subsubsec:method:spectra} the ANN model used by The Payne was validated by comparing synthetic spectra from the validation sample to spectra predicted by the ANN model using the same set of parameters.
Figure~\ref{Fig:histogram} shows the distribution of the median flux error using 3300 of the models from the validation set. It can be seen that the majority of the models have a median interpolation error just above 0.1~\%.
\FloatBarrier

\section{Effect on pseudo-continuum from metallicity}
\label{append:pseudocont}

\begin{figure*}
   \centering
   \includegraphics[width=\linewidth]{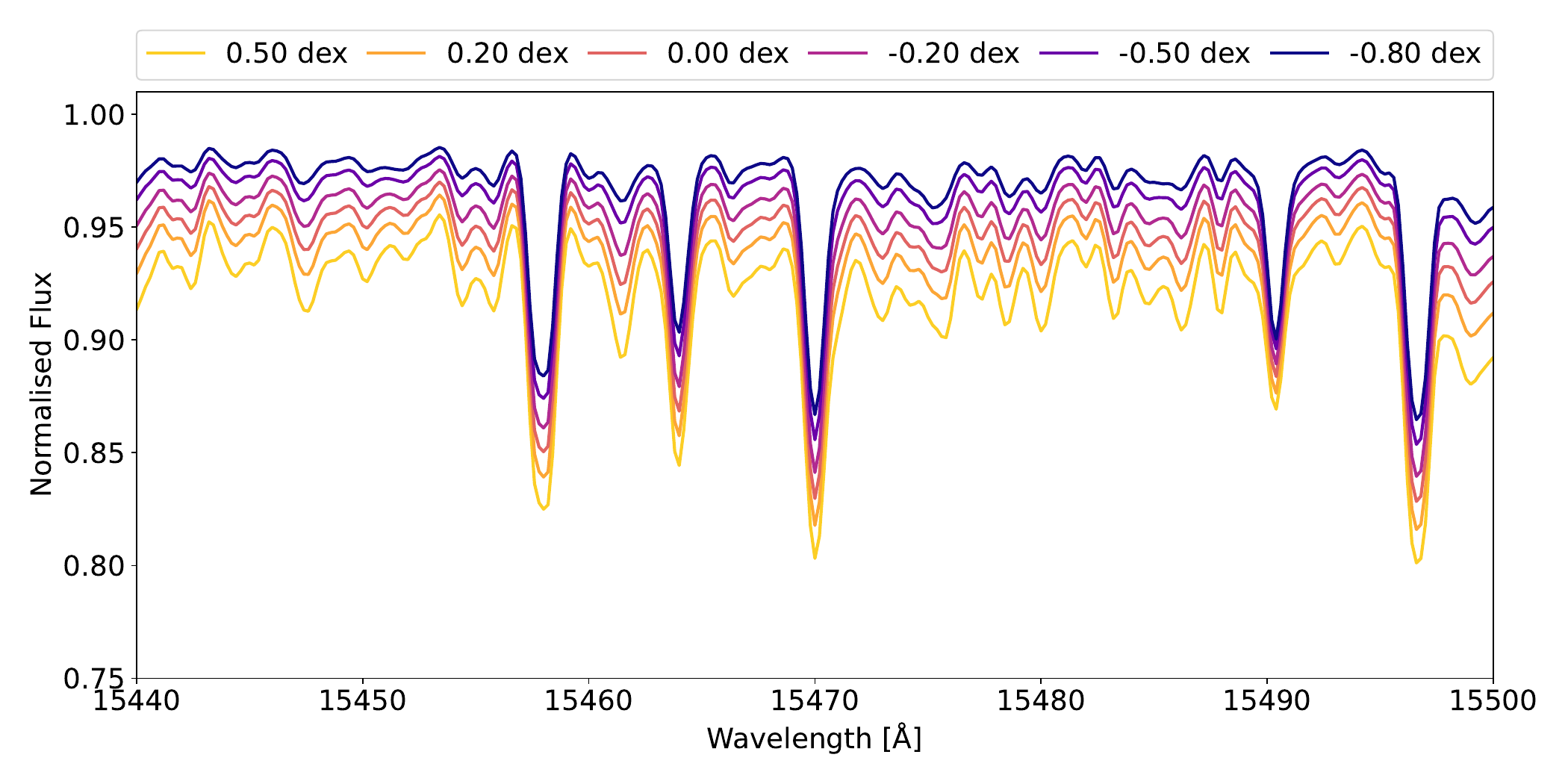}
   \caption{Example of H-band synthetic spectra generated with different metallicities. The $\teff$ and $\lgg$ values were set to 3800~K and 4.7~dex, respectively. The different colours correspond to different $\feh$ values.}
   \label{Fig:Water_cont_met_append}
\end{figure*}

Figure~\ref{Fig:Water_cont_met_append} shows synthetic spectra with varying metallicity, generated with Turbospectrum \citep{2022arXivGerber}, MARCS atmospheric models, the APOGEE DR16 line list \citep{APOGEE_DR16_linelist_Smith2021}, and the water line list by \citet{2018MNRASPolyansky}. The effective temperature was set to 3800~K and the surface gravity to 4.7~dex. The pseudo-continuum varies by about 0.03 continuum units between the lowest and highest metallicity.
\FloatBarrier

\section{Examples of other fitted spectra}
\label{append:fits}
\begin{figure*}
   \centering
   \hspace*{-2.5cm}
   \vspace*{-1cm}
   \includegraphics[width=23cm]{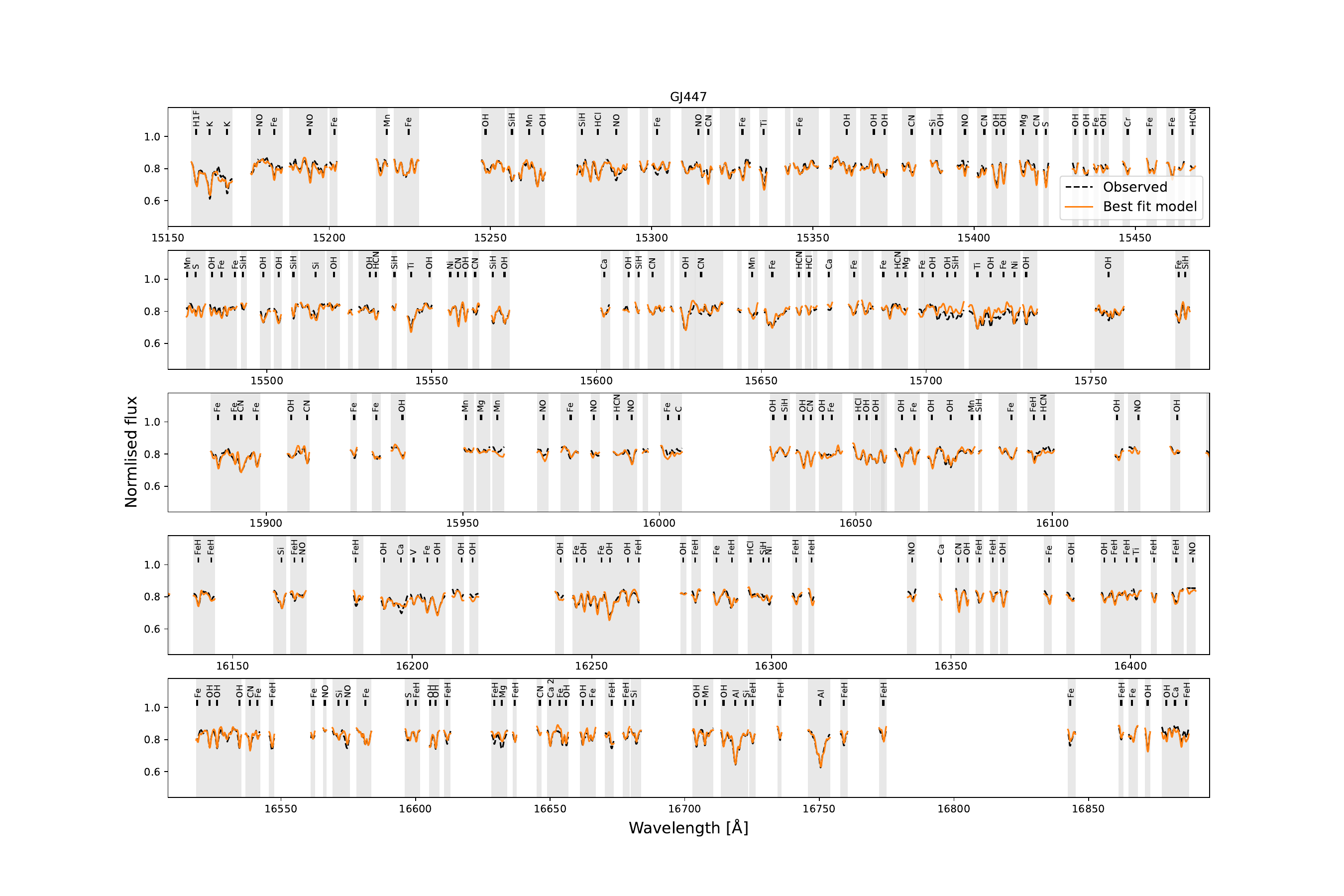}
   \caption{Normalised observed spectrum of the star GJ447 in dashed black and best-fit model in solid orange. Grey shaded areas indicate the location of the used line mask. The axes are on the same scale as in Fig.~\ref{Fig:GJ880_spec}. We obtained a $\teff$ of 3243~K, a $\lgg$ of 5.066 dex, and a metallicity of $-0.13$ dex.}
   \label{Fig:GJ447}
\end{figure*}

\begin{figure*}
   \centering
   \hspace*{-2.5cm}
   \vspace*{-1cm}
   \includegraphics[width=23cm]{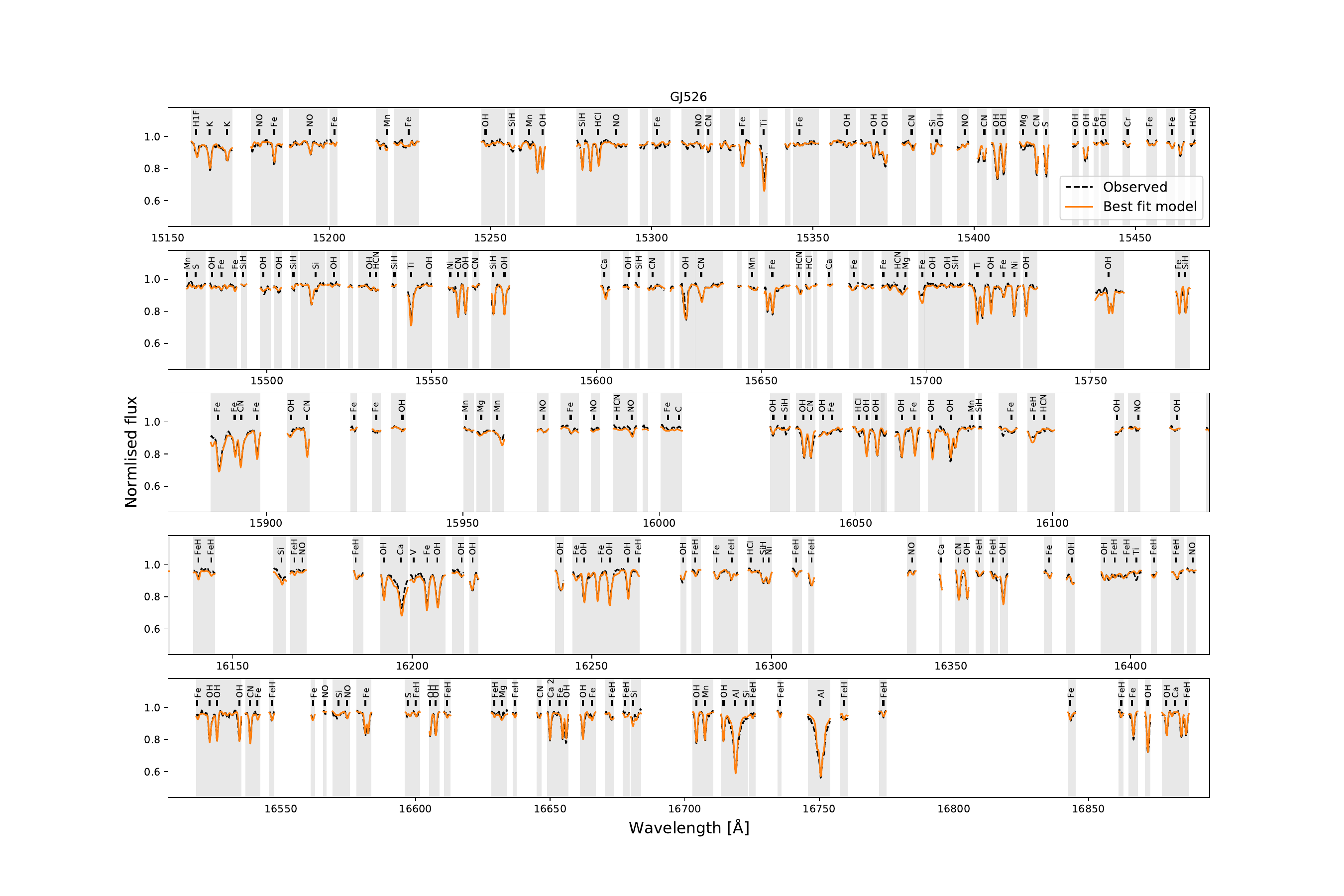}
   \caption{Same as Fig.~\ref{Fig:GJ447} for the star GJ526. We obtained a $\teff$ of 3729~K, a $\lgg$ of 4.792 dex, and a metallicity of $-0.32$ dex.}
   \label{Fig:GJ562}
\end{figure*}

\begin{figure*}
   \centering
   \hspace*{-2.5cm}
   \vspace*{-1cm}
   \includegraphics[width=23cm]{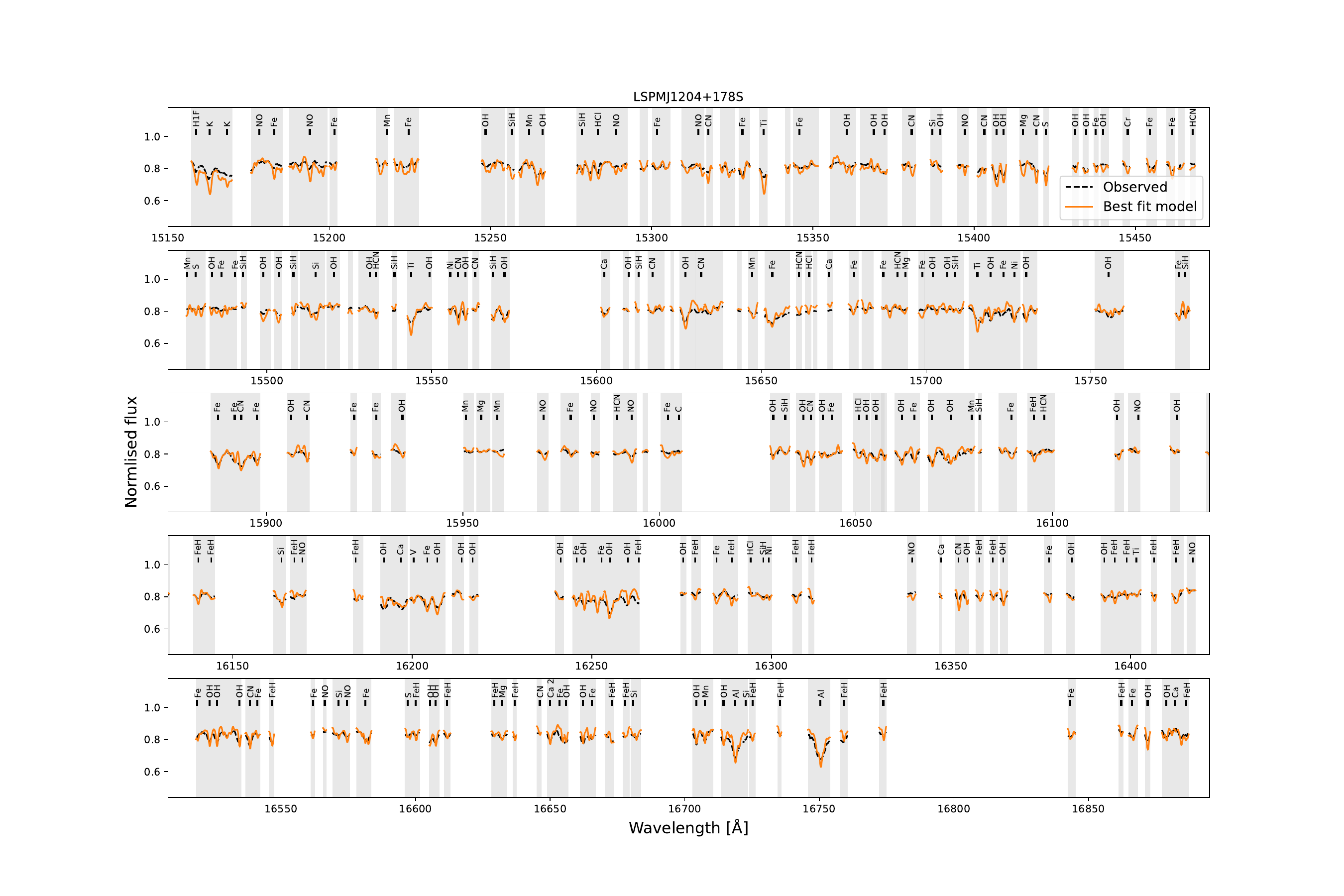}
   \caption{Same as Fig.~\ref{Fig:GJ447} for the fast rotating star LSPMJ1204+178S. Since the modified SAPP cannot fit for rotation the derived parameters are judged to be unreliable.}
   \label{Fig:fast_rotator}
\end{figure*}

Figures \ref{Fig:GJ447} and \ref{Fig:GJ562} shows spectra for the stars GJ~447 and GJ~526 and the corresponding best fit model obtained by the modified SAPP. The first star is a cooler M~dwarf where we derived a $\teff$ of 3243~K and the second star is warmer with a $\teff$ of 3729~K. For GJ447 we obtained a surface gravity with 5.066 dex and for GJ526 4.792 dex. Both stars have a sub-solar metallicity with $-0.13$ dex for GJ447 and $-0.32$ dex for GJ526. The regions outside of the line mask are not shown in the figures. 
For most of the lines the fit is good. It is slightly worse for the cooler star GJ447 especially at the redder region after 15700\AA. This could be due to missing molecular line data. We also can see discrepancies at the edges of the detector as was mentioned in Sect.~\ref{sec:Results}.
Figure \ref{Fig:fast_rotator} shows the spectra of the fast rotator mentioned in Sect.~\ref{sec:Results}. We can see in the figure that the fit is poor as the observed lines are wider than in the best-fit synthetic spectrum.  
\FloatBarrier

\end{appendix}

\end{document}